\documentclass[USenglish,oneside,twocolumn]{article}

\usepackage[utf8]{inputenc}
\usepackage[big]{dgruyter_NEW}

\usepackage{etex}
\usepackage{amssymb}
\usepackage{amsmath}

\usepackage{amsfonts}
\usepackage{color}
\usepackage{multirow}
\usepackage{extpfeil}
\usepackage[hyphenbreaks]{breakurl}
\usepackage{tikz}
\usetikzlibrary{calc,positioning,shapes.geometric,shadows}
\tikzset{multiple/.style = {double copy shadow={shadow xshift=1.8ex,shadow
         yshift=0.4ex,draw=black!30},fill=green!30,draw=black,thick,minimum height = 1cm,minimum width=2cm},}

\usepackage{version}
\includeversion{fullversion}\excludeversion{petsversion} 

\begin{petsversion}
\cclogo{\includegraphics{by-nc-nd.pdf}}
\end{petsversion}

\begin{document}

  \author[1]{Ghada Arfaoui}

  \author[2]{Jean-Fran\c{c}ois Lalande}

  \author[3]{Jacques Traor\'{e}}

  \author[4]{Nicolas Desmoulins}
  
  \author[5]{Pascal Berthom\'{e}}
  
  \author[6]{Sa\"id Gharout}

  \affil[1]{Orange Labs, F-14066 Caen, INSA Centre Val de Loire, Univ. Orléans, F-18020 Bourges, France, E-mail: ghada.arfaoui@orange.com}

  \affil[2]{INSA Centre Val de Loire - Inria, F-18020 Bourges, E-mail: jean-francois.lalande@insa-cvl.fr}
  
    \affil[3]{Orange Labs, F-14066 Caen, France, E-mail: jacques.traore@orange.com}

  \affil[4]{Orange Labs, F-14066 Caen, France, E-mail: nicolas.desmoulins@orange.com}

  \affil[5]{INSA Centre Val de Loire, F-18020 Bourges, E-mail: pascal.berthome@insa-cvl.fr}
  
  \affil[6]{Orange Labs, F-92130 Issy-les-moulineaux, France, E-mail: said.gharout@orange.com}

  \title{\huge A Practical Set-Membership Proof for Privacy-Preserving NFC Mobile Ticketing}

  \runningtitle{A Practical Set-Membership Proof for Privacy-Preserving NFC Mobile Ticketing}


  \begin{abstract}
{To ensure the privacy of users in transport systems, researchers are working on new protocols providing the best security guarantees while respecting functional requirements of transport operators. In this paper\footnote{\begin{fullversion}This paper is an extended version of the paper published at PETS 2015. \end{fullversion}The work  has been supported by the ANR-11-INS-0013 LYRICS Project.}, we design a secure NFC m-ticketing protocol for public transport that preserves users' anonymity and prevents transport operators from tracing their customers' trips. To this end, we introduce a new practical set-membership proof that does not require provers nor verifiers (but in a specific scenario for verifiers) to perform pairing computations. It is therefore particularly suitable for our (ticketing) setting where provers hold SIM/UICC cards that do not support such costly computations. We also propose several optimizations of Boneh-Boyen type signature schemes, which are of independent interest, increasing their performance and efficiency during NFC transactions. Our m-ticketing protocol offers greater flexibility compared to previous solutions as it enables the post-payment and the off-line validation of m-tickets. By implementing a prototype using a standard NFC SIM card, we show that it fulfils the stringent functional requirement imposed by transport operators whilst using strong security parameters. In particular, a validation can be completed in $184.25\,ms$ when the mobile is switched on, and in $266.52\,ms$ when the mobile is switched off or its battery is flat.}
\end{abstract}

\begin{petsversion}
  \journalname{Proceedings on Privacy Enhancing Technologies}
\DOI{Editor to enter DOI}
  \startpage{1}
  \received{2/15/2015}
  \revised{..}
  \accepted{5/15/2015}

  \journalyear{2015}
  \journalvolume{2015}
  \journalissue{2}
\end{petsversion}

\maketitle

\begin{fullversion}
\thispagestyle{empty}
\end{fullversion}

\section{Introduction}
\textit{Near Field Communication (NFC)}~\cite{iso14443} is a highly-practical emerging technology~\cite{nfcSTAT}. Indeed, NFC-enabled smartphones are being used in several domains, such as payment~\cite{mob-nfc}, access control~\cite{DST12} and ticketing~\cite{DKJ12}. In the following, we focus on mobile ticketing for public transport. Such a ticketing system is operated by a transport authority representing the transport service provider, and  it usually consists of three phases~\cite{nfc_public_transport}. First, the user registers (\textit{Registration}). Secondly, the user obtains the product (a set of m-tickets) to be used (\textit{Provisioning}). The last phase is the \textit{Validation}, during which an m-ticket is validated by the transport authority using the NFC connection of a mobile phone. During this phase, the m-ticketing system must ensure the integrity and authenticity of the m-ticket. Another important aspect is that a network connection~\cite{Chaumette2011} is sometimes required, either for the validator gate or for the smartphone itself. Moreover, the validation is subject to a hard time constraint imposed by the transport operators~\cite{gsma}: the m-ticket validation must occur in less than $300~ms$.

In the literature, many proposed solutions did not consider the security aspects. Even when security issues are addressed, the concept of user's \emph{privacy}, that is users' anonymity and users' trips unlinkability with respect to the transport service provider, is often overlooked~\cite{Ekberg2012,Tamrakar2013}. Some recent works~\cite{DKJ12,AFG13} have shown a special interest in users' privacy. Although they managed to ensure user's anonymity, these solutions suffer from privacy weaknesses.   

In this paper, we propose a new cryptographic protocol for m-ticketing that provides strong authentication, anonymity, and unlinkability properties, whilst remaining efficient when implemented in constrained environments such as SIM cards. 

Towards this goal, we also make various cryptographic optimizations. In particular, we present a new efficient set-membership proof enabling a prover to prove, in a zero-knowledge way, that his secret belongs to a given public set. Unlike previous constructions, this proof does not require pairing computations, especially on the prover side, which are quite costly, in the order of seconds, e.g. for the following microprocessors: $17.9s$ on an ATmega~\cite{SOS08}, $2.5s$ on a Philips HiPerSmart\texttrademark MIPS~\cite{DSD07}, $1.9s$ on an MSP430X~\cite{CLJ12}. We also propose several optimizations of Boneh-Boyen signature schemes. These results are of independent interest and increase the performance and efficiency of BB-signatures.

Based on these cryptographic primitives, our m-ticketing protocol enables to securely validate an m-ticket without disclosing any personal information to the transport operator, even if the latter is malicious. We push forward a strict privacy requirement when using one of the numbered m-tickets: for example, if the user uses the m-ticket number $3$ from the set $[1..10]$, the transport operator will only be able to check that the used m-ticket is one among the ten m-tickets $[1..10]$.

We implemented our system on a standard NFC SIM card. To the best of our knowledge, this is the first implementation of \textit{efficient} and  \textit{practical} set-membership proofs and BB-signatures on SIM cards. The entire validation process can be performed without any connection to a back-end server or computations delegation to the smartphone hosting the SIM card. This avoids any tracking of the user by a malware installed in the smartphone. We also show that the validation process occurs, on average, in $184.25ms$ when the mobile is switched on and in $266.52ms$ when the mobile is switched off or its battery is flat (battery-Off). Moreover, we show that our protocol supports the post payment approach (which provides more flexibility) and give countermeasures to detect any misuse of the service owing to the regular reports of unused m-tickets. 

The paper is structured as follows. Section~\ref{sec:related} discusses related work on privacy-preserving m-ticketing schemes. In Section~\ref{sec:framework}, we detail the framework of our m-ticketing protocol. Then, we introduce in Section~\ref{sec:req} our assumptions and the desired security, privacy and functional requirements that an m-ticketing system should fulfil. In Section~\ref{sec:cryp}, we describe the cryptographic assumptions and building blocks required for the sequel. We introduce our new set-membership proof in Section~\ref{sec:proof} and our m-ticketing protocol in Section~\ref{sec:sol}. Section~\ref{sec:analysis} gives the security proofs. In Section~\ref{sec:PoC}, we present our implementation before concluding in Section~\ref{sec:conclusion}.

\section{Related work}
\label{sec:related}
Heydt-Benjamin et al. were the first to propose a cryptographic framework for transport service~\cite{HCD06}.
They discuss the challenges that a privacy-preserving transit system should solve. 
Using cryptographic transit tickets should not disable basic capabilities like cloning detection, virtual ticket backups or transfers, ticket revocation. The efficiency of the use of a virtual ticket, especially over the air, is also important.
Later, many applied m-ticketing solutions have been proposed~\cite{Ekberg2012,Tamrakar2013,EAR13,AIC08,IVM12,DKJ12,AFG13}. However, each proposal has some limitations that this paper tries to solve. We briefly describe these limitations in this section.

First, the RSA based m-ticketing solution proposed by Ekberg and Tamrakar~\cite{Ekberg2012, Tamrakar2013}, the RFID e-tickets proposed by Sadeghi et al.~\cite{AIC08} and the ticketing solution proposed by Blass et al.~\cite{EAR13} only protect user's privacy with respect to outsiders and not the transport authority. The transport authority is considered honest which is not anymore a reasonable hypothesis.

Second,
the most recent protocols that include the expected level of privacy by protecting users' privacy even against the transport authority still have some efficiency problems. For instance, the BBS based protocol proposed by Isern-Deya et al.~\cite{IVM12} validates a ticket with a duration of few seconds, even if the implementation has been done on a smartphone. Some recent solutions were close to find the right balance between efficiency and users' privacy. Derler et al. proposal~\cite{DKJ12} provides anonymous tickets but users' trips are still linkable. Rupp et al. proposal~\cite{AFG13} provides a privacy preserving pre-payment protocol with refund for public transport with a refund verification procedure with a too high cost for constrained devices. Removing this verification makes possible to use the protocol, but it also enables a malicious transport authority to link users' trips\begin{fullversion}, as detailed in Appendix~\ref{appendix:flaw}\end{fullversion}\begin{petsversion}~\cite{ALT+15}\end{petsversion}.
   
Our m-ticketing protocol tries to fill the remaining gap between efficiency and strong privacy. It can be used in very constrained devices, like SIM cards and offers a high level of privacy. 

\section{Framework of the protocol}
\label{sec:framework}
In privacy-preserving m-ticketing system, we consider three different entities. The user (U) is the owner of an NFC-enabled smartphone and wants to use the transport service. The transport authority (TA) is the manager of the transport service. Additionally, we consider a third actor which is a revocation authority (RA). It can revoke m-tickets and users' anonymity, and is completely independent of the transport authority. This role may be split between several authorities in such a way that they should all agree and cooperate before recovering the identity of an m-ticket holder. 

An m-ticketing system consists of six different phases in our protocol. (1) The m-ticketing system parameters and keys are initialized during the \emph{initialization}. (2) The \emph{registration} phase enables a user to register to the transport service. (3) In the \emph{permission token request} phase, a user gets a \textit{permission token} allowing him to generate $max_{ticket}$ m-tickets. (4) The \emph{validation} phase consists in generating and validating an m-ticket. (5) The \emph{revocation} phase enables to retrieve the identity of a user who generated a given m-ticket and the m-tickets obtained by a given user. Finally, (6) the \emph{reporting} phase enables a user to report his usage (i.e. the m-tickets that he did not use) to the transport authority that can detect any duplication of m-tickets (i.e. m-tickets that have been validated several times). In the sequel, these phases are modeled with various algorithms and protocols executed by the above entities. 

\textbf{Initialization.} 

$\mathtt{Setup(1^{\lambda})}$: This probabilistic algorithm outputs $pp$ a description of the system parameters. We assume that $pp$ are implicit to the other algorithms, and that they include $\lambda$, the security parameter, and $max_{ticket}$, the number of m-tickets that each book/set of m-tickets contains. They are also an implicit input to the adversary, we will then omit them. 

$\mathtt{Keygen}$({\small $pp$}): This probabilistic algorithm outputs the two following secret/public key pairs: $(rsk, rpk)$ for the revocation authority and $(tsk, tpk)$ for the transport authority. The system public key $gpk$ is eventually set as ($pp$, $tpk$, $rpk$).

\textbf{Registration.}  

$\mathtt{UKeygen}$({\small $gpk$, $ID_U$}): This probabilistic algorithm outputs a secret/public key pair $(usk, upk)$ for a user identifier $ID_U$.

$\mathtt{Register}$({\small $U(ID_U, upk), TA(DB_{REG}$})): This is an interactive protocol between a new user that takes as input his identity $ID_U$ and his public key $upk$, and the TA that takes as input the database $DB_{REG}$ where the identifiers of the registered users will be stored. If TA accepts the protocol, the user's identity and public key are stored within $DB_{REG}$. 

\textbf{Permission token request.} 

$\mathtt{TokenRequest}$({\small U$(upk, gpk)$, TA$(tsk, gpk, DB_{REG})$}): This is an interactive protocol between a user that takes as input $(upk,$ $gpk)$, and the TA that takes as input $(tsk, gpk, DB_{REG})$. If the user accepts the protocol, his output is a permission token $\tau$ that will enable him to generate/validate a set of $max_{ticket}$ m-tickets. If TA accepts the protocol, its output is a transcript \emph{view} of the protocol. 

\textbf{Validation.} 

$\mathtt{GenTicket}$({\small $gpk, \tau$}): This probabilistic algorithm takes as input a user's permission token $\tau$ and outputs an m-ticket $Tick_k$ with a serial number $B_k$ such that $k \in [1..max_{ticket}]$.

$\mathtt{ValidateTicket}$({\small $gpk, Tick_k$}): This deterministic algorithm  takes as input a ticket $Tick_k$. If $Tick_k$ is valid, it outputs $1$ (otherwise $0$) and $Tick_k$ is stored within  the database $DB_{UsedTickets}$ that will be used to detect the m-tickets that have been used several times.

\textbf{Revocation.} 

$\mathtt{IdentUser}$({\small $rsk$, $DB_{REG}$, $Tick_k$}): This deterministic algorithm takes as input the private key $rsk$, the database $DB_{REG}$ and a valid m-ticket $Tick_k$. It outputs the identifier $ID_U$ of the user who obtained $Tick_k$. If $ID_U$ does not belong to $DB_{REG}$, it outputs $\perp$.

$\mathtt{IdentTicket}$({\small $rsk$, \emph{view}, $ID_U$}): This deterministic algorithm takes as input the private key $rsk$, a user's identifier $ID_U$ and a transcript \emph{view} of an execution of $\mathtt{TokenRequest}$ with this user. It outputs all the m-tickets that can be generated from the token obtained after the execution of the $\mathtt{TokenRequest}$ protocol that led to \emph{view}.

\textbf{Reporting.} 

$\mathtt{ReportTicket}(\tau)$: This algorithm is executed by a user with his permission token $\tau$. The user generates all the unused m-tickets and collects them in a \textit{usage report} $R$. $R$ is then sent to the transport authority.

$\mathtt{IdentDuplicate}$({\small $B_k, DB_{UsedTickets}$}): This deterministic algorithm takes as input $B_k$ the serial number of a valid m-ticket $Tick_k$ and $DB_{UsedTickets}$, and outputs the number of occurrences of $B_k$ in $DB_{UsedTickets}$.

\section{Requirements}
\label{sec:req}
In this section, we first describe our trust assumptions. Then, we detail the requirements of our m-ticketing system. We consider two types of requirements, \emph{functional} requirements, which are ``efficiency'' and ``versatility'' and \emph{security and privacy} requirements, which consist in ``correctness'', ``unforgeability'', ``unlinkability'' and ``non-frameability''.

\subsection{Trust assumptions}
At the user's side, we consider an untrusted mobile platform (smartphone) with an incorporated trusted secure element. We use a SIM card as a tamper resistant secure element since it could be certified EAL4+~\cite{SAT10}. Using a SIM card will ensure the integrity of the implementation. Indeed, the m-ticketing cardlet manages the cryptographic credentials of the m-ticketing application and executes required cryptographic computations without delegating any of them to the smartphone. Consequently, if the mobile application is compromised, the security will not be impacted because all the sensitive data and credentials are stored and managed in the SIM card. If the SIM card gets compromised, we planned countermeasures (cf. Section~\ref{subsub:couter}).

\subsection{Functional requirements}
\subsubsection{Efficiency} M-ticketing systems must fulfil functional requirements imposed by transport operators~\cite{gsma}, in particular the validation of an m-ticket must be performed in less than $300ms$. We must consider that a SIM card has limited computation capabilities. In particular, pairing APIs are not available on current SIM cards.
\subsubsection{Versatility} 
The mobile phone and the validator cannot be assumed to be connected to a back-end server during the validation phase. This enables the user to use an m-ticket in any kind of situation, especially in areas with low connectivity, e.g., underground, or if the battery of his mobile is flat. Moreover, the m-ticketing system must support the post-payment mode, i.e., charged later (after its use).  
\subsection{Security and privacy model}
\label{sub:secmodel}
Besides the correctness property\footnote{\scriptsize Informally speaking, our protocol is correct if (1) a valid permission token enables to generate valid m-tickets, (2) honestly generated m-tickets are accepted, (3) a validated m-ticket enables revocation authorities to identify the user who generated it and (4) revocation authorities can retrieve all the m-tickets generated by a given registered user that obtained a valid permission token.} (which is obvious), we formally define three security and privacy properties of our m-ticketing protocol in which the attack capabilities of a probabilistic polynomial time adversary $\mathcal{A}$ are modeled by providing him access to some oracles. In the sequel, $\mathcal{HU}$ will denote the set of honest users and $\mathcal{MU}$ the set of corrupted users. We assume that $\mathcal{A}$ receives all the exchanged messages in our system. $\mathcal{A}$ acts as an active adversary as regards to the messages issued by malicious users and as a passive adversary with respect to honest users.

$\mathcal{O}Register_{HU}$ is an oracle that will be used by an adversary in order to register honest users. By calling this oracle with $ID_U$ as argument, the adversary adds a new user. The oracle runs $(upk, usk) \leftarrow \mathtt{UKeygen}(gpk, ID_U)$ and adds $ID_U$ (along with $upk$) to the set $\mathcal{HU}$. The private key $usk$ is kept secret and public key $upk$ is returned to the adversary.

$\mathcal{O}Register_{MU}$ is an oracle that will be used by an adversary in order to register malicious users. The adversary calls this oracle with argument the identifier $ID_U$ of a user and sets his public key to $upk$ and his private key to $usk$. The identity $ID_U$ (along with $upk$) is added to the set $\mathcal{MU}$.

$\mathcal{O}CorruptUser$ is a user secret key oracle enabling the adversary to obtain the private key $usk$ of a user  $ID_U \in \mathcal{HU}$. The oracle transfers $ID_U$ to $\mathcal{MU}$ and returns $usk$. 

$\mathcal{O}TokenRequest_U$ is an oracle that runs the user's side in the $\mathtt{TokenRequest}$ protocol. This oracle will be used by an adversary playing the role of a malicious TA. The adversary gives to the oracle an identity $ID_U$ of an honest user and his public key $upk$. The adversary is then given a transcript view of the protocol. 

$\mathcal{O}TokenRequest_T$ is an oracle that runs the transport authority side in the $\mathtt{TokenRequest}$ protocol. This oracle will be used to simulate the execution of the protocol between a user (corrupted or not) and an honest TA.

$\mathcal{O}GenTicket(ID_U, \emph{view})$ is an oracle that takes as input the identifier $ID_U$ of an honest user and a transcript \emph{view} of an execution of the $\mathtt{TokenRequest}$ protocol with this user and outputs an m-ticket $Tick_k$  using a fresh index $k$ that has not been used in a previous query of $\mathcal{O}GenTicket$ on $ID_U$ and \emph{view}. The oracle records ($ID_U$, $Tick_k$) in a list \emph{Set}. 

$\mathcal{O}IdentTicket_T(ID_U, \emph{view})$ is an oracle that takes as input the identifier of a user $ID_U$ and a transcript \emph{view} of an execution of $\mathtt{TokenRequest}$ with this user and outputs all the m-tickets that this user is able to generate.

$\mathcal{O}IdentUser_T(Tick_k)$  is an oracle that returns the identifier $ID_U$ of the user who generated an m-ticket $Tick_k$.

$\mathcal{O}ReportTicket(ID_U, \emph{view})$ is an oracle that takes as input the identifier $ID_U$ of an honest user and a transcript  \emph{view} of a $\mathtt{TokenRequest}$ execution with this user and outputs the set of unused m-tickets.  For each unused m-ticket $Tick_{k}$, the oracle records $(ID_U, Tick_k)$ in \emph{Set}.

In the sequel, we denote by $\mathcal{A}$(\emph{keys}, \emph{DB}: \emph{oracles}) an adversary who receives the keys ``\emph{keys}''. This adversary has only \emph{read access} to the databases ``\emph{DB}''  and can query the oracles ``\emph{oracles}''.

\subsubsection{Non-frameability}
Informally speaking, it should be impossible for anyone to falsely accuse an honest user of having spent an m-ticket. We formally define the non-frameability experiment $\mathbf{Exp}_{\mathcal{A}}^{Nfra}(1^\lambda)$ in Figure~\ref{fig:nonfra}. The scheme is non-frameable, if for any probabilistic polynomial time adversary $\mathcal{A}$, the probability\\ $\mathtt{Pr}$[$\mathbf{Exp}_{\mathcal{A}}^{Nfra}(1^\lambda)$=1] is negligible.

\begin{figure}[tb]
\scriptsize
\begin{tabular}{l}
\parbox{\linewidth}{
\normalfont
\mathversion{normal}
$\mathbf{Exp}_{\mathcal{A}}^{Nfra}(1^\lambda)$
\begin{enumerate}\itemsep1.75pt \parskip0pt \parsep0pt
            	\item $pp \leftarrow Setup(1^{\lambda});~\mathcal{HU} \leftarrow \varnothing;~\mathcal{MU} \leftarrow \varnothing;~Set \leftarrow \varnothing$.  
            	\item $(gpk, tsk, rsk) \leftarrow \texttt{Keygen}(pp)$.
				\item $(Tick_k) \leftarrow \mathcal{A}$({$gpk$, $tsk$, $rsk$, $DB_{REG}$, $DB_{UsedTickets}$: $\mathcal{O}Register_{HU}$, $\mathcal{O}Register_{MU}$, $\mathcal{O}CorruptUser$, $\mathcal{O}TokenRequest_U$, $\mathcal{O}GenTicket$, $\mathcal{O}ReportTicket$}).
				\item If $\texttt{ValidateTicket}(gpk, Tick_k)=0$ or $\texttt{IdentUser}(rsk,$ $DB_{REG}, Tick_k) = \perp$ then return 0.
				\item If $\texttt{IdentUser}(rsk, DB_{REG}, Tick_k) = ID_U \in \mathcal{HU}$  and $(ID_U, Tick_k) \notin Set$ then return 1 else return 0.
            \end{enumerate}\vspace{-0.2cm}%
}\\
\end{tabular}
\caption{Non-frameability security experiment}
\label{fig:nonfra}
\end{figure}

\subsubsection{Unforgeability}
Informally speaking, it should be impossible for anyone (1) to validate more m-tickets than what he obtained i.e. an adversary who retrieved $N$ tokens $\tau$ ($N$ sets of $max_{ticket}$ m-tickets) should not be able to generate more that $N*max_{ticket}$ m-tickets; (2) to validate m-tickets such that the algorithm $\texttt{IdentUser}$ returns $\perp$, i.e., an identifier $ID_U$ that doesn't appear in $DB_{REG}$. We formally define the unforgeability experiment $\mathbf{Exp}_{\mathcal{A}}^{unforg}(1^\lambda)$ in Figure~\ref{fig:unf}. The scheme is unforgeable if for any probabilistic polynomial time adversary $\mathcal{A}$, the probability $\texttt{Pr}[\mathbf{Exp}_{\mathcal{A}}^{unforg}(1^\lambda)=1]$ is negligible.

\begin{figure}[tb]
\scriptsize
\begin{tabular}{l}
\parbox{\linewidth}{
\normalfont
\mathversion{normal}
$\mathbf{Exp}_{\mathcal{A}}^{unforg}(1^\lambda)$
\begin{enumerate}\itemsep1.75pt \parskip0pt \parsep0pt
\item $pp \leftarrow Setup(1^{\lambda});~\mathcal{HU} \leftarrow \varnothing;~\mathcal{MU} \leftarrow \varnothing$.
\item $(gpk, tsk, rsk) \leftarrow \texttt{Keygen}(pp)$.
\item $(\{Tick^j_{k_j}\}^{j=l}_{j=1}, \{R_i\}^{i=f}_{i=1}) \leftarrow \mathcal{A}$({$gpk$: $\mathcal{O}Register_{HU}$, $\mathcal{O}Register_{MU}$, $\mathcal{O}CorruptUser$, $\mathcal{O}TokenRequest_T$, $\mathcal{O}GenTicket$, $\mathcal{O}ReportTicket$}). An $R_i$ corresponds to a ``usage report'', i.e. a set of unused m-tickets.
\item Let $DB$ be an empty database.
\item For $j$ from $1$ to $l$ do~\{If $\texttt{ValidateTicket}(gpk, Tick^j_{k_j})$ then store $Tick^j_{k_j}$ in $DB$\}. 
\item For $i$ from $1$ to $f$ do~\{Validate the m-tickets of the report $R_i$ and store valid unused m-tickets in $DB$\}.
\item For all $Tick_k$ in $DB$ do \{$b=\texttt{IdentDuplicate}(B_k, DB)$ where $B_k$ is the serial number of the m-ticket $Tick_k$\\
If b>1, then delete all the duplicates of the m-ticket $Tick_k$.\\
If $\texttt{IdentUser}(rsk, DB_{REG}, Tick_k)$ outputs $\perp$ then return $1$ and aborts.\}.
\item If $L$, the number of m-tickets that remained within $DB$, is greater that $N*max_{ticket}$ ($L > N*max_{ticket}$) where $N$ is the number of calls of the oracle {$\mathcal{O}TokenRequest_T$} and $max_{ticket}$ is the number of authorized m-tickets by token, then return $1$ else return $0$.
\end{enumerate}\vspace{-0.2cm}%
}\\
\end{tabular}
\caption{Unforgeability security experiment}
\label{fig:unf}
\end{figure}

\subsubsection{Unlinkability}
\label{unlink}
Informally speaking, it should be impossible, except for the revocation authorities, to trace the m-tickets obtained by a user, in particular: (1) to link m-tickets obtained during the permission token request phase to the validated/used ones; (2) to link two m-tickets validated by the same user or to decide whether two validated m-tickets have the same number/index or not; (3) to link validated m-tickets to non-used m-tickets reported by the user to the transport authority. For this, an adversary has full control over the transport authority (in particular it owns the private key $tsk$) and all the users except two honest users $i_0$ and $i_1$. The adversary can initiate the $\texttt{IdentUser}$ protocol over any m-ticket and can get the user's identity behind it, except for the m-tickets generated by $i_0$ and $i_1$. He can also initiate the $\texttt{IdentTicket}$ protocol for all the users except for $i_0$ and $i_1$. We define the unlinkability experiment $\mathbf{Exp}_{\mathcal{A}}^{unlink}(1^\lambda)$ in Figure~\ref{fig:unlink}. The scheme is unlinkable if for any probabilistic polynomial time adversary $\mathcal{A}$, the advantage $\mathbf{Adv}_{\mathcal{A}}^{unlink-b}(1^\lambda) = |Pr[\mathbf{Exp}_{\mathcal{A}}^{unlink-b}(1^\lambda) = b]- 1/2|$ is negligible. 

\begin{figure}[tb]
\scriptsize
\begin{tabular}{l}
\parbox{\linewidth}{
\normalfont
\mathversion{normal}
$\mathbf{Exp}_{\mathcal{A}}^{unlink-b}(1^\lambda)$
\begin{enumerate} \itemsep1.75pt \parskip0pt \parsep0pt
\item $pp \leftarrow Setup(1^{\lambda});~\mathcal{HU} \leftarrow \varnothing;~\mathcal{MU} \leftarrow \varnothing$.
\item $(gpk, tsk, rsk) \leftarrow \texttt{Keygen}(pp)$.
\item ($i_0$, $k_0$, $i_1$, $k_1$) $\leftarrow \mathcal{A}$({$gpk$, $tsk$, $DB_{REG}$, $DB_{UsedTickets}$:  $\mathcal{O}Register_{HU}$, $\mathcal{O}Register_{MU}$, $\mathcal{O}CorruptUser$, $\mathcal{O}TokenRequest_U$, $\mathcal{O}GenTicket$, $\mathcal{O}IdentTicket_T$, $\mathcal{O}IdentUser_T$, $\mathcal{O}ReportTicket$)}.
\item If $i_0 \in \mathcal{MU}$ or $i_1 \in \mathcal{MU}$ then output $\perp$.
\item
	\begin{itemize}
		\item[(a)] let $i_0$ and $i_1$ run the protocol $\texttt{TokenRequest}$ and get the permission tokens $\tau_0$ and $\tau_1$ and output $\emph{view}_0$ and $\emph{view}_1$. 
		\item[(b)] $Tick_{k_b} \leftarrow \texttt{GenTicket}(gpk, \tau_0)$ and\\  $Tick_{k_{1-b}} \leftarrow \texttt{GenTicket}(gpk, \tau_1)$, with $b \in \{0,1\}$.
	\end{itemize}
\item $b' \leftarrow \mathcal{A}$({$gpk$, $tsk$, $DB_{REG}$, $DB_{UsedTickets}$, $Tick_{k_j}$, $Tick_{k_{1-j}}$: $\mathcal{O}Register_{HU}$, $\mathcal{O}Register_{MU}$, $\mathcal{O}CorruptUser$, $\mathcal{O}TokenRequest_T$, $\mathcal{O}GenTicket$, $\mathcal{O}IdentTicket_T$, $\mathcal{O}IdentUser_T$, $\mathcal{O}Report$ $Ticket$}), with $j \in \{0,1\}$.
\item If {$\mathcal{O}CorruptUser$ was requested  on $i_0$ or $i_1$, or $\mathcal{O}IdentTicket_T$} was requested on ($i_0$, $\emph{view}_0$) or ($i_1$, $\emph{view}_1$) then output $\perp$. 
\item If {$\mathcal{O}IdentUser_T$} was requested for $Tick_{k_j}$ or $Tick_{k_{1-j}}$, output $\perp$.
\item If {$\mathcal{O}ReportTicket$} was requested for $i_0$ or $i_1$ and $i_0$ and $i_1$ did not validate the same number of m-tickets then output $\perp$.
\item Return $b'$.
\end{enumerate}\vspace{-0.2cm}%
}\\
\end{tabular}
\caption{Unlinkability security experiment}
\label{fig:unlink}
\end{figure}

\textbf{Remark:} another basic requirement that our ticketing system should fulfill is the \textit{traceability} meaning that revocation should succeed in the face of attacks by malicious users. In fact, our formulations of \textit{unforgeability} and \textit{non-frameability} capture this requirement.
Indeed, an attacker who could produce a ticket which either (1) cannot be traced to a user or (2) identify an honest user who didn't obtain this ticket, would either break the \textit{unforgeability} requirement (1) or the \textit{non-frameability} requirement (2).

\section{Cryptographic assumptions and building blocks}
\label{sec:cryp}
In this section, we introduce the notations, definitions and cryptographic tools used in the description of our set-membership proof and m-ticketing protocol. 

\subsection{Bilinear Maps}
We consider throughout this document, except when it is explicitly mentioned, bilinear maps $e : G_1 \times G_2 \rightarrow G_T$ where all groups $G_1$, $G_2$ and $G_T$ are multiplicative and of prime order $p$. The mapping $e$ satisfies the following properties:
\begin{eqnarray*} 
\forall g_1 \in G_1, \  g_2  \in G_2\ and\ a,b \in \mathbb{Z}_p, \  e(g_1^a,g_2^b ) = e(g_1,g_2)^{ab}\\
For\ g_1 \neq 1_{G_1 }\ and\ g_2 \neq 1_{G_2 },\ e(g_1,g_2) \neq 1_{G_T }\\
e\ \text{is efficiently computable}.
\end{eqnarray*}

\subsection{Computational assumptions}
\textbf{Decisional Diffie-Hellman (DDH).}
For any multiplicative group $G$ of prime order $p$, the DDH assumption states that given a random generator $g \in G$, two random elements $g^a$, $g^b$ in $G$, and a candidate $X \in G$, it is hard to decide whether $X =g^{ab}$ or not.

\textbf{eXternal Diffie-Hellman (XDH).}
Given three groups $G_1$, $G_2$, $G_T$, and a bilinear map $e : G_1 \times G_2 \rightarrow G_T$, the XDH assumption states that DDH assumption holds in $G_1$.

\textbf{q-Strong Diffie-Hellman (q-SDH).}
The q-SDH assumption holds for some group $G_1$ if it is hard, given ($g$, $g^y$, $g^{y^2},\ldots,g^{y^q }$) $\in$ $G_1^{q+1}$, to output a pair $(x,g^{1/(y+x)})$.

\textbf{q-Decisional Diffie-Hellman Inversion (q-DDHI).}
In any multiplicative group $G$ of prime order $p$, the q-Decisional Diffie-Hellman Inversion assumption states that, given a random generator $g \in G$ and the values $(g, g^{\alpha}, g^{\alpha^2},\ldots, g^{\alpha^q}) \in G$, for a random $\alpha \in \mathbb{Z}_p$ and a candidate $X \in G$, it is hard to decide whether $X=g^{1/\alpha}$ or not. 

\textbf{The Decisional Composite Residuosity Assumption.} There is no probabilistic polynomial time distinguisher for $n$-th residues modulo $n^2$. In other words, there is no probabilistic polynomial time adversary that can distinguish $\mathbb{S}$ from $\mathbb{Z}^{*}_{n^2}$, where $\mathbb{S} = \{z \in \mathbb{Z}^{*}_{n^2},$ $\exists y \in \mathbb{Z}^{*}_{n^2}: z = y^n mod\ n^2\}$.

\subsection{Zero-knowledge proof of knowledge} 
Roughly speaking, a zero knowledge proof of knowledge, denoted ZKPK, is an interactive protocol during which a prover convinces a verifier that he knows a set of secret values ($\alpha_1, \alpha_2,\ldots, \alpha_n$) verifying a given relation $\Re$ without revealing anything else. Such a proof will be denoted in the sequel $POK((\alpha_1, \alpha_2,\ldots, \alpha_n): \Re(\alpha_1, \alpha_2,\ldots, \alpha_n))$. A ZKPK should be \textbf{complete} (a valid prover is accepted with overwhelming probability), \textbf{sound} (a false prover should be rejected with overwhelming probability) and \textbf{zero-knowledge} (no information about the secret is revealed). These ZKPK can be made non-interactive using the generic transformation introduced by Fiat and Shamir~\cite{FS87}. The resulting non-interactive proofs, sometimes called \textit{signatures of knowledge}~\cite{CPS96}, can be proven secure in the random oracle model of~\cite{BR93} (see~\cite{PS96}). Such a signature of knowledge, for the relation $\Re$, will be denoted $SOK((\alpha_1, \alpha_2,\ldots, \alpha_n): \Re(\alpha_1, \alpha_2,\ldots, \alpha_n))$  .

\begin{fullversion}

\subsection{Camenisch-Lysyanskaya signatures}
These signatures, proposed by Camesnich and Lysyanskaya~\cite{CL04}, are equipped with additional protocols. One of these protocols allows a signature to be issued on messages that are not known by the signer, but for which the signer only knows a commitment. Informally, in a protocol for signing a committed value, we have a signer with public key $pk$, and the corresponding secret key $sk$, and a user who queries the signer for a signature. The common input to the protocol is a commitment $C$, known by both parties, on secret values $(x_1, x_2,\ldots,x_n)$ known only by the user. At the end of this protocol, the user obtains a valid $CL-signature = Sign(x_1, x_2,\ldots,x_n)$ and the signer learns nothing about $(x_1, x_2,\ldots,x_n)$.

Another protocol allows to prove knowledge of a signature on a tuple of messages $(x_1, x_2,\ldots,x_n)$ without
releasing any information on the corresponding signature. Each message can either be revealed to the verifier, sent in a committed form, or it may be such that the verifier has no information on it. In particular, it is possible to prove the knowledge of a CL-signature on committed values.

\end{fullversion}

\subsection{Set Membership Proofs}
A set membership proof allows a prover to prove, in a zero-knowledge way, that his secret lies in a given public set. Such proofs can be used, for instance, in the context of electronic voting, where the voter needs to prove that his secret vote belongs to the set of all possible candidates. Recent propositions of set membership proofs~\cite{CLZ12,CCJ13} follow the same idea: a designated authority produces public signatures on each element of the public set $\Phi$ (and only on these elements). The proof of knowledge of a secret $x \in \Phi$ consists in proving the knowledge of a (public) signature on the secret $x$ (which will be only possible if $x$ belongs to the set $\Phi$) without revealing $x$ or the used signature. Solutions in~\cite{CLZ12,CCJ13} require the prover to perform pairing computations. However, these cryptographic tools are not often convenient for many constrained platforms.

\subsection{Boneh-Boyen Signatures}
The signature scheme used by the designated authority (which could be the verifier) to sign each element of the set $\Phi$ is the one proposed by Boneh and Boyen~\cite{BBS04,BB04,BB08}. Based on their scheme, which is secure under the q-SDH assumption, it is possible to prove knowledge of a signature on a message, without revealing the signature nor the message. 

\subsubsection{Boneh-Boyen Signatures with pairings.}
Having a secret key $y$ and two random generators $g_1$, $g_2$ of $G_1$, the signature of a message $m \in \mathbb{Z}_p$ is obtained by computing $\sigma = g_{1}^{1/(y+m)}$. Given a bilinear pairing $e$, a signature $\sigma$ of $m$ is valid if $e(\sigma, Y g_2^m ) = e(g_1, g_2)$, where $Y = g_2^y$ is the signer's public key.  

\subsubsection{Boneh-Boyen Signatures without pairings.} 
Let $G$ be a cyclic group with prime order $p$ where the Decisional Diffie-Hellman (DDH) problem is assumed to be hard and $g_1$, $g_2$ two random generators of $G$. The signer's private key is $y \in \mathbb{Z}_p$ and its public key is $Y=g_{2}^y$.

Similarly to Boneh-Boyen signatures with pairings, the signature on a message $m$ is the value $A =g_{1}^{1/(y+m)}$. This implies that $A^y = g_1 A^{-m}$. Since we work in a group $G$ not equipped with a bilinear map, the signer must additionally prove that the signature on $m$ is valid, which is done by generating a ZKPK $\pi$ that the discrete logarithm of ($g_1 A^{-m}$) in the base $A$ is equal to the discrete logarithm of $Y$ in the base $g_2$: $SOK(y: Y = g_2^y \wedge A^y = g_1 A^{-m})$. Such a proof of equality of discrete logarithms has been introduced in~\cite{CP92}. Finally, the signature $A$ on $m$ is valid if the proof $\pi$ is valid.

\textbf{Theorem 1.} 
The Boneh-Boyen (BB for short) signature scheme without pairings is existentially unforgeable under a weak chosen message attack under the q-Strong Diffie-Hellman assumption, in the random oracle model.

\textbf{Proof (sketch).} Under the q-Strong Diffie-Hellman assumption, it is impossible to find a message $m$ (which was not given to the signing oracle) and a value $A$ such that $A=g_1^{1/(y+m)}$, as proved in~\cite{BB04,BB08}. Moreover, in the random oracle model, the proof of knowledge $\pi$ is unforgeable~\cite{BB04,BB08}, which concludes the proof.

\newcommand{\mymulticolumn}[3]{\multicolumn{#1}{>{\normalfont\mathversion{normal}\raggedright}#2}{#3}}
\begin{figure*}
\scriptsize
\centering
\begin{tabular}{>{\normalfont\mathversion{normal}\raggedright}p{\dimexpr 0.4\linewidth-\tabcolsep}
>{\normalfont\mathversion{normal}\raggedright}p{\dimexpr 0.1\linewidth-2\tabcolsep}
>{\normalfont\mathversion{normal}\raggedright\arraybackslash}p{\dimexpr 0.5\linewidth-\tabcolsep}}
\mymulticolumn{1}{c}{\textbf{Prover}} & &\mymulticolumn{1}{c}{\textbf{Verifier}\rule[-3pt]{0pt}{12pt}}\\ \hline
\textbf{Public input:} public parameters, sets $\Phi$ and $\sum$ & & \textbf{Public Input:} public parameters, sets $\Phi$ and $\sum$\\
\textbf{Private Input:} $k$  & & \textbf{Private Input:} $y$\\
\hline
\mymulticolumn{3}{c}{\textbf{Pre-computations:~~~~~~~~~~~~~~~~~~~~~~~~~}}\\ \hline
\mymulticolumn{3}{l}{\textbf{Choose} $\nu \in \mathbb{Z}_p^*$ and \textbf{Compute:} $Com=g_1^k h_T^{\nu}$}\\
\mymulticolumn{3}{l}{\textbf{Pick} the valid BB-signature corresponding to the element $k:$ $A_k=g^{1/(y+k)}$}\\
\mymulticolumn{3}{l}{\textbf{Choose} $l \in \mathbb{Z}_p^*$ and \textbf{Compute (see remark 1):} $B = A_k^l$; $B_1 = B^{-1}$; $D = B_1^k g^l$}\\
\mymulticolumn{3}{l}{\textbf{Choose} $k_1,l_1,r_1 \in \mathbb{Z}_p^*$ and \textbf{Compute:} $Com_1 = g_1^{k_1} h_T^{r_1}$ and $D_1 = B_1^{k_1} g^{l_1}$}\\
\hline
\mymulticolumn{3}{c}{\textbf{Real time computations:~~~~~~~~~~~~~~~~~~~~~~~~~}}\\
\hline
\mymulticolumn{1}{l}{\textbf{Compute:} $c = H(Com, B, D, Com_1, D_1, ch)$}&$\xtwoheadleftarrow{~~~~ch~~~~}$& \textbf{Choose} $ch \in \mathbb{Z}_p^*$ (A random challenge) \\
\mymulticolumn{1}{l}{$s_1 = k_1 + c \times k~mod~p$; $s_2= r_1 + c \times v~mod~p$;}&&\\
\mymulticolumn{1}{l}{$s_3= l_1 + c \times l~mod~p$; }&&\tikz[overlay,remember picture]{\node[anchor=base] (U) {}; }\textbf{Check} that $B \neq 1_{G_1}$\\
\mymulticolumn{1}{l}{$\Pi=Com, B, D, s_1, s_2, s_3$  \tikz[overlay,remember picture]{\node[anchor=base] (R) {}; 
\draw[->>,color=black,line width=0.25pt] (R.east) .. controls +(2,0) and +(-2,0) .. node[above]{$\Pi$} (U.north west); }}  &&$\bullet$ If the verifier and the designated entity are the same entity, it implies that the verifier holds the private signature key $y$ \textbf{(First case)}. Hence the prover does not send the value $D$. The verifier can compute it: $D = B^y$ and goes to (*)\\
&&$\bullet$ Otherwise, if the verifier doesn't know the private signature key $y$ \textbf{(Second case)}, then, it checks that $e(D,g_3) = e(B,Y)$ and goes to (*)\\
&& \textbf{(*) Compute:}  $\tilde{C} = g_1^{s_1} h_T^{s_2} Com^{-c}$ and $\tilde{D} = B_1^{s_1} g^{s_3} D^{-c}$ \\
&& \textbf{Check} that: $c = H(Com, B, D, \tilde{C}, \tilde{D}, ch)$ \\
\end{tabular}
\caption{A new efficient set membership proof} 
\label{tab:proof}
\end{figure*}

\subsection{Threshold Cryptosystems}
The El Gamal cryptosystem~\cite{ElGamal85} works as follows. Let $G$ be a cyclic group with prime order $p$ where the Decisional Diffie-Hellman (DDH) problem is assumed to be hard. The public key consists of the elements $(g_T, h_T = g_T^{x_T})$, where $g_T$ is a random generator of $G$, and the corresponding private key is formed by $x_T \in \mathbb{Z}_p^*$. The El Gamal ciphertext of a message $m \in G$ is $(C_1 = g_T^r, C_2 = m \times h_T^r )$, where $r \in \mathbb{Z}_p^*$ is a random number. The decryption of the ciphertext $(C_1, C_2)$ is obtained through the following computation: $m = C_2/C_1^{x_T} $. The El Gamal cryptosystem is semantically secure under the DDH assumption. 

The Paillier cryptosystem~\cite{PAI99} works as follows. Let $a$ and $b$ be random primes for which $a, b>2, a \neq b, |a| = |b|$ and $gcd(ab,(a-1)(b-1)) = 1$. Let $n = ab$, $\Pi = lcm(a-1, b-1)$, $K = \Pi^{-1}~mod~n$, and $g_P = (1+n)$. The public key is $pk = (n, g_P)$ and the secret key is $sk = (a, b)$. To encrypt $m \in \mathbb{Z}_n$, the user chooses $r \in \mathbb{Z}^*_n$ and $C = EncryptPai_{pk}(m, r) = g_P^m r^n~mod~n^2$. The decryption algorithm is given by $DecryptPai_{sk}(C) = ((C^{\Pi K}mod~n^2)-1) /n~mod~n^2 = m$. The Paillier cryptosystem is secure under the Decisional Composite Residuosity assumption\footnote{\scriptsize We use Paillier encryption scheme in our system in order to satisfy the unforgeability requirement even in a concurrent setting, where the adversary is allowed to interact with the transport authority, during the permission token request phase, in an arbitrarily interleaving (concurrent) manner (see~\cite{EJ07}).}.

In a threshold version, the El Gamal public key and its corresponding private key~\cite{DF89} (resp. the Paillier public key and its corresponding private key~\cite{PJ01}) are cooperatively generated by $n$ parties; however, the private key is shared among the parties. In order to decrypt a ciphertext, a minimal number of $t$ (the threshold) out of $n$ parties is necessary.

\subsection{Pseudo-random Function}
A useful building block of our m-ticketing protocol is the pseudo-random function introduced by Dodis and Yampolskiy~\cite{DY05,DY03}. Their construction works as follows. Let $G$ be a cyclic group with prime order $p$, $g_t$ a generator of $G$, $s \in \mathbb{Z}_p$ a private secret. The pseudo-random function (PRF for short) $F_s$ takes as input a message $k \in \mathbb{Z}_p$ and outputs $F_s(k)= g_t^{1/(s+k+1)}$. This construction is secure under the q-DDHI assumption in $G$.

\section{A new set membership proof}
\label{sec:proof}
In this section, we present a new set membership protocol. Our set membership proof is a ZKPK which allows the prover to convince any verifier that the committed integer $k$ lies in the public discrete set $\Phi$. This ZKPK can be made non-interactive using the Fiat-Shamir heuristic and the random oracle model~\cite{FS87}. The commitment scheme that we use is the one proposed by Pedersen~\cite{Pedersen92}. 

Our set membership proof bears some similarities with the protocol proposed by Camenisch et al.~\cite{CCS08}. Yet, our set membership proof does not ask the prover (a SIM card and/or a smartphone in our setting) nor the verifier (in a specific scenario) to perform pairing computations. Our set membership proof is also more efficient and conceptually simpler than the recent one proposed by Canard et al. at EuroPKI 2013~\cite{CCJ13}. Their scheme which involves several verifiers, who share a decryption key, seems to be tailored to specific applications such as e-voting where the verifiers (tallying authorities) own such a key to open ballots.  
 
Let $G_1$, $G_2$ and $G_T$ be three (multiplicative) bilinear groups of prime order $p$ and a bilinear map $e : G_1 \times G_2 \rightarrow G_T$. Let $g,  g_1$ and $h_T$ be three generators of $G_1$ and consider a generator $g_3$ of $G_2$. A designated authority (which may or may not be the verifier) randomly chooses a value $y \in \mathbb{Z}_p^*$ (its private key) and publishes the corresponding public key $Y = g_3^y$. After generating its key, this designated authority can issue BB-signatures on each element of the public set $\Phi$. Let $k$ denote an element of the set $\Phi$ and $A_k$ the BB-signature on the value $k$, i.e., $A_k=g^{1/(y+k)}$.  The set of all message-signature pairs $(k, A_k)$, that we denote by $\sum$, is published by the designated authority. $H$ will denote a hash function, for instance, SHA-256.

In Figure~\ref{tab:proof}, we propose a protocol enabling to prove that the value $k$ committed by the prover in the commitment $Com$ belongs to the set $\Phi$ which comes down to proving knowledge of the BB-signature $A_k$.

\textbf{Remark 1:}
Proving knowledge of a valid BB-signature $A_k=g^{1/(y+k)}$ on a value $k$ committed in $Com=g_1^k h_T^{\nu}$, without revealing neither $k$ nor the signature and without using pairings on the prover's side can be done as follows.
The prover first randomizes $A_k$ by choosing a random value $l \in \mathbb{Z}_p^*$ and computes $B=A_k^l$. Since $A_k=g^{1/(y+k)}$ this implies that $B=A_k^l=(g^l )^{1/(y+k)}$ and then that 
$B^{y+k}=g^l$. It implies, if we note $B_1=B^{-1}$ and $D=B^y$ that $D=B_1^k g^l$.

As a result, in order to prove that the value $k$ committed in $Com$ belongs to the set $\Phi$, the prover just has to send $B$ and $D$ to the verifier and compute the following ZKPK: $\Pi = POK(k,\nu,l: Com=g_1^k h_T^\nu \wedge D=B_1^k g^l)$. The computation of the ZKPK $\Pi$ is described in Figure~\ref{tab:proof}.

Upon receiving $\Pi$, the verifier proceeds to the verification. We distinguish two cases. In the \emph{first} case, the verifier and the designated authority are the same. Thus, the verifier holds the private signature key $y$. Consequently, the verification of $\Pi$ occurs without requiring any pairing computations. This implies that our set membership proof does not require pairing computations for either the prover or the verifier. In the \emph{second} case, the verifier does not have the private signature key $y$ of the designated authority. Then, the verifier needs  to perform some pairing computations in order to verify $\Pi$. Nevertheless, the prover still has no pairing computations to perform. This is particularly interesting for the m-ticketing use case.

\textbf{Theorem 2.} If the $|\Phi|-$SDH assumption holds, then the protocol in Figure~\ref{tab:proof} is a zero-knowledge argument of set membership for a set $\Phi$. The proof is in Appendix~\ref{appendix:proof2}.
 
\section{Our m-ticketing protocol}
\label{sec:sol}
This section begins with an overview of our protocol which main notations are summarized in Table~\ref{tab:notation}, before moving to our m-ticketing protocol and its post-payment aspects. 
 
\begin{table}%
\scriptsize%
\centering%
\noindent%
\begin{tabular}{p{\dimexpr 0.27\linewidth-2\tabcolsep}
p{\dimexpr 0.73\linewidth}}
$G_1$,$G_2$,$G_T$ & Multiplicative groups of prime order p \\
$g$,$g_0$,$g_1$,$g_t$,\\$g_T$,$g_U$,$h$,$G$,$H$ & Generators of $G_1$\\ 
$(x_T, h_T)$ & ElGamal private and public keys of the revocation authorities \\ 
$((a,b), (n, p_P))$ & Pailler private and public keys of the revocation authorities \\
$(\gamma, (W, W'))$& BB-signature scheme private and public keys of the transport authority\\ 
$(y, Y)$& BB-signature private and public keys of the transport authority for set membership proof\\
$(x_U, h_U)$& Private and public key of the user\\
$A$ & A permission token\\
$ID_U$ & The identity of the user\\
$Tick_k$ & The $k^{th}$ m-ticket\\
$B_k$ & The serial number of the $k^{th}$ m-ticket\\
$E$ & The ElGamal encryption of the user secret corresponding to his set of m-tickets\\
$\Pi$ & A proof (POK)\\
$ch$ & A challenge \\
$DB_{REG}$ & The TA secure registration database\\
$DB_{UsedTickets}$ & A secure database where the TA centralises the used m-tickets\\
\end{tabular}
\caption{Notation used in the m-ticketing protocol}
\label{tab:notation}
\end{table}

\subsection{Overview}
\label{sub:overview} 
We assume that m-tickets have a unique rate per area such as in Paris~\cite{paris}, Berlin~\cite{berlin} and Moscow~\cite{moscow} undergrounds. The user can have different types of m-tickets sets such that every set is characterized by a rate, an area and the number of available m-tickets ($max_{ticket}$), e.g., a set of 10 m-tickets valid in area 1 and the price of one m-ticket is 1.30\texteuro.

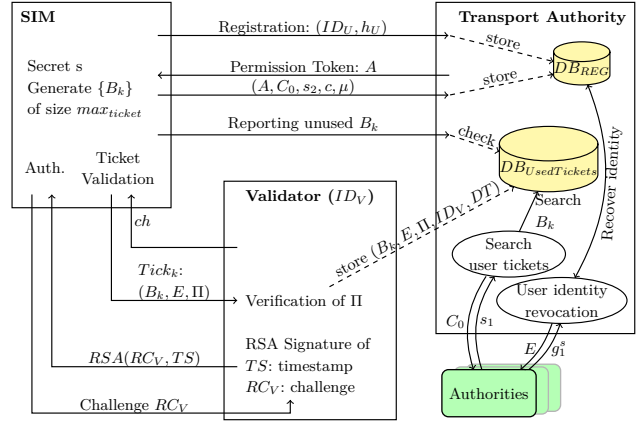
\begin{figure}[!t]%
\centering%
\scalebox{0.62}{%
\begin{tikzpicture}[y=0.80pt,x=0.80pt,inner sep=0pt, outer sep=0pt,   line width=0.7pt, database/.style={
      cylinder,
      cylinder uses custom fill,
      cylinder body fill=yellow!40,
      cylinder end fill=yellow!40,
      shape border rotate=90,
      aspect=0.35,
      draw
    },    decision/.style={ 
        draw,
        rectangle, rounded corners
    }]
\begin{scope}
  \path[draw=black,line join=miter,line cap=butt, rounded
    corners=0.0000cm] (30,475) rectangle (150,320);
  \path[fill=black, inner sep=5] (30,475) node[below right,font=\bfseries] (SIM) {SIM};
  
  \node at (85,410) [align=left] {Secret s\\ Generate \{$B_k$\}\\ of size $max_{ticket}$};

  \path[draw=black,line join=miter,line cap=butt,rounded
    corners=0.0000cm] (350,475) rectangle (500,225);
  \path[fill=black, inner sep=5] (500,475) node[below left,font=\bfseries] (TA) {Transport Authority};

  \path[draw=black,line join=miter,line cap=butt,rounded
    corners=0.0000cm] (190,340) rectangle (320,160);
	\path[fill=black, inner sep=5] (255,340) node[below,font=\bfseries] (Validator) {Validator ($ID_V$)};

	\node[database] (dbreg) at (460,425) {$DB_{REG}$};
	\node[database] (dbusedtickets) at (435,350) {$DB_{UsedTickets}$};

   \node (register) at (360,450) {};
   \draw[->] (140,450) -- (register) node[above,midway]  {Registration: $(ID_U, h_U)$};
   \draw[->, dashed] (register) -- (dbreg) node[above,midway,sloped,inner sep=2pt] {store};
   
   \draw[<-] (140,420) -- (360,420)  node[above,midway,inner sep=2pt] {Permission Token: $A$};
    \node (register2) at (360,405) {};
    \draw[->] (140,405) -- (register2)  node[above,midway] {$(A, C_0	, s_2, c, \mu)$};
    \draw[->, dashed] (register2) -- (dbreg) node[above,midway,sloped,inner sep=2pt] {store}; 
    
    \draw[->] (140,375) -- (360,375)  node[above,midway,inner sep=2pt] {Reporting unused $B_k$};
    \node (reporting) at (360,375) {};
    \draw[->, dashed] (reporting) -- (dbusedtickets) node[above,midway,sloped,inner sep=2pt] {check};

\node[right,inner sep=5pt] (tickK) at (200,250) {Verification of $\Pi$};
\draw[<-] (120,330) -- node[right,midway,inner sep=2pt] {$ch$} (120,290) --  (200,290);
\draw[->] (105,330) -- (105,250) -- node[above,midway,align=left] {$Tick_k$:\\ $(B_k, E, \Pi)$} (tickK);
\draw[->, dashed] (tickK) -- (dbusedtickets) node[midway, above, sloped, align=center] {store $(B_k,E,\Pi, ID_V, DT)$};
\node at (110, 350) [align=center] {Ticket \\ Validation};

\node[right,inner sep=5pt,align=left] (RSA) at (200,200) {RSA Signature of \\ $TS$: timestamp \\ $RC_V$: challenge};
\draw[<-] (60,330) --  (60,200) -- node[above,midway] {$RSA(RC_V, TS)$}  (RSA) ;
\draw[->] (45,330) -- (45,165) -- node[above,midway] {Challenge $RC_V$} (200,165)  -- (240,165) -- (240,175);
\node at (55,350) {Auth.};

\node[decision,multiple] (auth) at (390,180)  {Authorities};

\node[draw, ellipse] (revocation) at (445,250) [align=center] {User identity \\ revocation};
\draw[<-] (auth) edge [bend left=-15] node[left,midway,inner sep=2pt] {$E$} (revocation) ;
\draw[->] (auth) edge [bend left=-15,transform canvas={xshift=5}] node[right,midway,inner sep=2pt] {$g_1^s$} (revocation) ;
\draw[<->] (dbreg) edge [bend left=25] node[sloped,below] {Recover identity} (revocation) ;
\end{scope}

\node[draw, ellipse] (used) at (405,285) [align=center] {Search \\ user tickets};
\draw[<-] (auth) edge [bend left=25,transform canvas={xshift=-5}] node[left,midway,inner sep=2pt] {$C_0$} (used) ;
\draw[->] (auth) edge [bend left=25]node[right,midway,inner sep=2pt] {$s_1$} (used) ;
\draw[->] (used) -- (dbusedtickets) node[midway,right,align=left,inner sep=4pt] {Search \\ $B_k$};
\end{tikzpicture}%
}%
\caption{The m-ticketing protocol overview}%
\label{tab:overview}%
\end{figure}

As shown in Figure~\ref{tab:overview}, at the beginning, the user registers at the transport service. Then, he retrieves a permission token $A$ from the transport authority.  This token enables the generation of $max_{ticket}$ m-tickets (an m-ticket set). The token $A$ is a BB-signature on a secret $s$ known only by the user. The secret $s$ identifies a type of m-tickets set. 

At the validation phase, the user authenticates the validator and sends an m-ticket $Tick_k$: $(B_k, E, \Pi)$. $B_k$ is a unique serial number of the m-ticket, $E$ is an ElGamal encryption of $g_1^s$, and $\Pi$ is a ZKPK. $\Pi$ proves that (1) the user has a valid BB-signature on $s$ without revealing neither $s$ nor the signature $A$, (2) $E$ is an encryption of $g_1^s$ without revealing $s$ and (3) $k$ belongs to $[1..max_{ticket}]$ without revealing $k$. (3) is based on the new set membership proof of Section~\ref{sec:proof}. 

Finally, the user post-pays by regularly reporting unused m-tickets i.e. by revealing the unused m-ticket serial numbers $B_k$ to the TA. This enables to detect any malicious behaviour, i.e., validating more m-tickets than what were obtained, without breaking the user's privacy. Additional countermeasures allows to recover the user's identity and retrieve his m-tickets, with the agreement of the authorities.

\subsection{Model}
\textbf{Setup of public parameters.}
Let $g$, $g_0$, $g_1$, $g_t$, $g_T$, $g_U$, $h$, $G$, $H$ be nine generators of $G_1$ and $g_2$, $g_3$ two generators of $G_2$. The user's SIM card will perform computations only in $G_1$ (note that current SIM cards are not designed to handle pairing computations nor computations in $G_2$ or $G_T$).

\textbf{Revocation authorities.}
The revocation authorities set two pairs of keys: a private and public keys of the threshold ElGamal cryptosystem $(pk_{RG}, sk_{RG})$ and a private and public keys of the threshold Paillier cryptosystem $(pk_{RP}, sk_{RP})$. 

We assume that the private keys are shared among the revocation authorities (e.g. using the technique proposed in~\cite{GJK03} for El Gamal and in~\cite{PJ01} for Paillier). In order to decrypt a ciphertext, a minimal number $t$ (the threshold) of these authorities should cooperate.

The public key, $pk_{RG}$, consists of the elements $(g_T, h_T = g_T^{x_T})$, where $g_T$ is a random generator of $G_1$, and the corresponding private key, $sk_{RG}$, is formed by $x_T \in \mathbb{Z}_p^*$. 
Let $a$ and $b$ be random primes for which $a, b > 2, a \neq b, |a| = |b|$ and $gcd(ab,(a-1)(b-1)) = 1$; let $n = ab$, $\Pi = lcm(a-1, b-1)$, $K = \Pi^{-1}~mod~n$, and $g_P = (1+n)$; then the public key, $pk_{RP}$, is $(n, g_P)$ and the secret key, $sk_{RP}$, is $(a, b)$.

\textbf{Transport authority.}
The transport authority (TA) sets for each group, i.e. a type of m-tickets set, a key pair of the BB-signature scheme. The private key is a random element $\gamma \in \mathbb{Z}_p^*$ and the corresponding public key is  $W'= g_0^\gamma$ and $W=g_2^\gamma$.
The transport authority sets another key pair of the BB-signature scheme which will be used for the ``Set membership proof''. The private key is a random element $y \in \mathbb{Z}_p^*$ and the corresponding public key is  $Y = g_3^y$.

\textbf{User.}
The user owns a public/private key pair $(x_U \in \mathbb{Z}_p^*,h_U=g_U^{x_U})$. During the permission token request, the user obtains from TA a token $A =(g_1^s h)^{1/(\gamma+r)}$. $s$ is jointly chosen by TA and the user but is only known by the user (whereas both know $r$).

\begin{figure*}
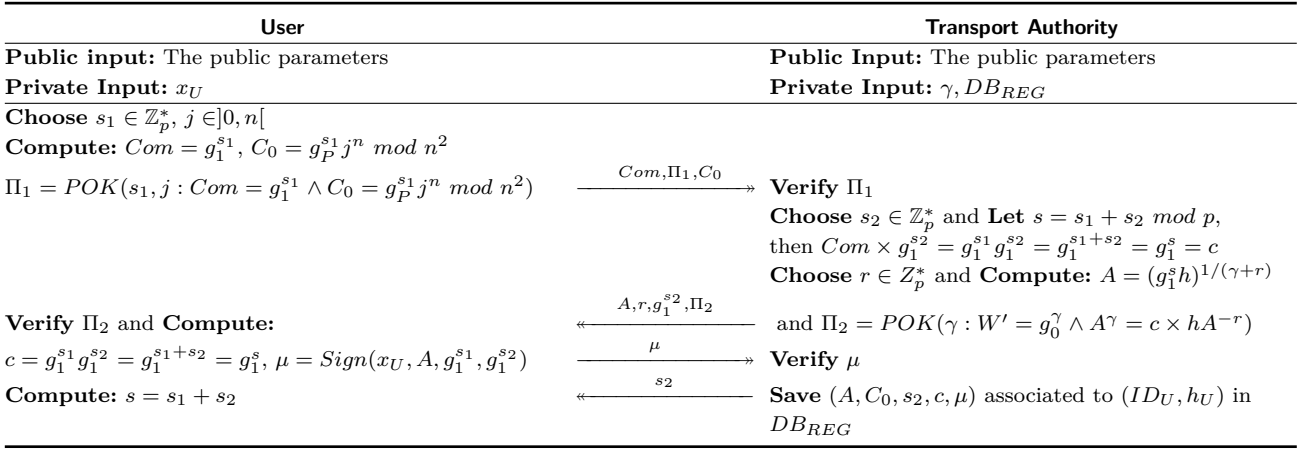

\scriptsize
\centering
\begin{tabular}{>{\normalfont\mathversion{normal}\raggedright}p{\dimexpr 0.43\linewidth-\tabcolsep}
>{\normalfont\mathversion{normal}\raggedright}p{\dimexpr 0.15\linewidth-2\tabcolsep}
>{\normalfont\mathversion{normal}\raggedright\arraybackslash}p{\dimexpr 0.42\linewidth-\tabcolsep}}
\multicolumn{1}{c}{\textbf{User}} & &\multicolumn{1}{c}{\textbf{Transport Authority} \rule[-3pt]{0pt}{12pt}} \\ \hline
\textbf{Public input:} The public parameters & & \textbf{Public Input:} The public parameters\\
\textbf{Private Input:} $x_{U}$ & & \textbf{Private Input:} $\gamma, DB_{REG}$\\
\hline
\textbf{Choose} $s_1 \in \mathbb{Z}^*_p$, $j \in ]0, n[$ &&\\ 
\textbf{Compute:} $Com=g_1^{s_1}$, $C_0 = g_{P}^{s_1} j^n~mod~n^2$ &&\\
$\Pi_1=POK(s_1, j: Com=g_1^{s_1} \wedge C_0 = g_{P}^{s_1} j^n~mod~n^2)$& $\xtwoheadrightarrow{~~~Com,\Pi_1,C_0~}$&\textbf{Verify} $\Pi_1$\\
&&\textbf{Choose} $s_2 \in \mathbb{Z}^*_p$ and \textbf{Let} $s=s_1+s_2~mod~p$,\\
&& then $Com \times g_1^{s_2}= g_1^{s_1} g_1^{s_2}=g_1^{s_1+s_2}=g_1^s=c$ \\
&& \textbf{Choose} $r \in Z_{p}^{*}$  and \textbf{Compute:} $A = (g_1^{s} h)^{1/(\gamma+r)}$\\
\textbf{Verify} $\Pi_2$ and \textbf{Compute:}&$\xtwoheadleftarrow{~~A, r, g_1^{s_2}, \Pi_2~~~}$& \vspace{-0.30cm} and $\Pi_2=POK(\gamma: W'=g_0^\gamma \wedge A^{\gamma}=c \times h A^{-r})$\\
 $c=g_1^{s_1} g_1^{s_2}=g_1^{s_1+s_2}=g_1^s$, $\mu = Sign(x_U, A, g_1^{s_1}, g_1^{s_2})$&$\xtwoheadrightarrow{~~~~~~~\mu~~~~~~~~~}$&\textbf{Verify} $\mu$\\
\textbf{Compute:} $s=s_1+s_2$&$\xtwoheadleftarrow{~~~~~~~s_2~~~~~~~~}$&\textbf{Save} $(A, C_0, s_2, c, \mu)$ associated to $(ID_U, h_U)$ in $DB_{REG}$\\
\end{tabular}
\caption{The protocol of the permission token request}
\label{tab:permissionTokenRequest}
\end{figure*}

\subsection{M-ticketing protocol}
Our protocol is divided in three phases: user registration, permission token request and validation. The later includes the validator authentication and the m-ticket validation.

\subsubsection{User registration phase}
We denote by $ID_{U}$ a user's identity and $DB_{REG}$ the database where the TA saves the identities of registered users. First, the user sends his public key $h_U$ and his identity $ID_{U}$ to TA. Then, he signs, using Schnorr signature scheme~\cite{CPS91}, a random challenge $rc$ received from TA, using his private key $x_U$. If the signature is valid, TA saves the tuple $(ID_{U}, h_U)$ in $DB_{REG}$. Then, the user securely introduces his banking credentials in order to be subsequently charged.

\subsubsection{Permission token request phase}
The permission token request phase is detailed in Figure~\ref{tab:permissionTokenRequest}. A permission token $(A = (g_1^{s} h)^{1/(\gamma+r)}, r)$ consists of an (extended) BB-signature~\cite{CL04} on $s$ (the secret member group key only known by the user, whereas $r$ is known by the user and TA).  Thanks to his permission token, the user will be able to use $max_{ticket}$ m-tickets. The value of $max_{ticket}$ is set by TA and linked to the key pair $(\gamma,W = g_2^\gamma)$ used by TA during the permission token request protocol. TA will use a different pair of keys for each possible value of $max_{ticket}$, i.e., for each group associated to each type of m-tickets set.

At the end of this phase, TA saves in $DB_{REG}$ the \emph{view} $(A, C_0, s_2, c,$ $\mu)$\, where $C_0$ is the Paillier encryption of $s_1$ such that the secret $s= s_1+s_2$ and $(c,\mu)$ are the commitment and signature by the user of his secret $s$.

\subsubsection{Validation phase}
In this phase, the validator is authenticated, the permission token is verified and finally the m-ticket is validated.

\smallskip
\textbf{Validator authentication.}
The validator authentication consists of a challenge / response protocol. The user sends a Random Challenge $RC_{V}$  to the validator gate. Upon receiving $RC_{V}$, the validator replies with the RSA signature on a timestamp ($TS$) and the received challenge (we use a short public verification exponent $v$ in order to have a fast verification in the SIM card). 
Then, the m-ticketing cardlet checks the received signature. If it succeeds, the m-ticketing cardlet checks the validity of the permission token based on the timestamp ($TS$) sent by the validator. Indeed, if $TS$ is lower than $\mathcal{D}$ ($TS<\mathcal{D}$), the permission token is still valid. Then, the cardlet checks whether the number of used m-tickets reached $max_{ticket}$, the number of authorized post-paid m-tickets.  If a check fails, the m-ticketing cardlet aborts and the m-ticketing application displays a message to ask the user to renew the permission token.

\smallskip
\textbf{M-ticket validation.}
An m-ticket $Tick_k$ (indexed $k$) is characterized by a serial number $B_k=g_t^{1/(s+k+1)}$ along with an ElGamal encryption $E=(C_1 =g_T^a,C_2=g_1^s \times h_T^a )$ of $g_1^s$. To prove the validity of a m-ticket, the user must prove that:
\begin{itemize}
\itemsep-0em 
\item $k$ belongs to $\Phi=[1..max_{ticket}]$  using our new set membership proof (described in Section~\ref{sec:proof}),
\item He knows a valid BB-signature $A=(g_1^s h)^{1/(\gamma+r)}$  on $s$ (without revealing both s and the signature),
\item $E$ is an encryption of $g_1^s$.
\end{itemize}
Therefore $Tick_k=(B_k,E,\Pi)$ where $\Pi = POK(k,a,s,r,A: B_k=g_t^{1/(s+k+1)} \wedge A=(g_1^s h)^{1/(\gamma+r)} \wedge C_1 =g_T^a \wedge C_2=g_1^s \times h_T^a \wedge k \in [1..max_{ticket}])$.

\begin{figure*}
\scriptsize
\centering
\begin{tabular}{>{\normalfont\mathversion{normal}\raggedright}p{\dimexpr 0.40\linewidth-\tabcolsep}
>{\normalfont\mathversion{normal}\raggedright}p{\dimexpr 0.14\linewidth-2\tabcolsep}
>{\normalfont\mathversion{normal}\raggedright\arraybackslash}p{\dimexpr 0.44\linewidth-\tabcolsep}}
\mymulticolumn{2}{c}{\textbf{M-ticketing cardlet}}  &\mymulticolumn{1}{c}{\textbf{Validator} \rule[-3pt]{0pt}{12pt}}\\ \hline
\textbf{Public input:} The public parameters and the public keys of the revocation and transport authorities $h_T, W, Y$. The sets $\Phi=[1..max_{ticket}]$ and $\sum=\{A_1, A_2,\ldots,A_{max_{ticket}}\}$ where $A_i=g^{1/(y+i)}$ for $i \in \Phi$&& \textbf{Public Input:} The public parameters, $h_T, W, Y$. The sets $\Phi$ and $\sum$ \\ 
\textbf{Private Input:} $A=(g_1^s h)^{1/(\gamma+r)}$, $k$ (the index of the ticket that will be used), $s$& & \textbf{Private Input:} The private signature keys $\gamma$ and $y$ (in some scenario)\\ 
$\mathcal{D}$ day: validity end date of the permission token&&\\
\hline
\mymulticolumn{3}{c}{\textbf{Pre-computations:}}\\ \hline
\mymulticolumn{1}{l}{\textbf{Compute} $B_k=g_t^{1/(s+k+1)}$} &&\\
\mymulticolumn{3}{l}{\textbf{Choose} $a \in \mathbb{Z}^{*}_p$ and \textbf{Compute:} $C_1=g_T^a, C_2=g_1^s \times h_T^a$}\\
&&\\
\mymulticolumn{3}{l}{\underline{\textit{Computation of elements involved in the proof that the user knows a valid signature on $s$}}}\\
\mymulticolumn{3}{l}{\textbf{Choose} $\alpha \in \mathbb{Z}^{*}_p$ and \textbf{Compute (see remark 2):} $B_0=A^{\alpha}$; $B'=B_0^{-1}$; $C=g_1^{\alpha s} \times h^{\alpha} \times B'^{r}$}\\
\mymulticolumn{3}{l}{\textbf{Choose} $r_2, r_3, r_4, a_1, d_1, b_1, \alpha_1, t, t_1, t_3, t_4, t_5, \nu,$ $d', f' \in  \mathbb{Z}^{*}_p$}\\
\mymulticolumn{3}{l}{\textbf{Compute:} $T'=G^s H^{r_2}$; $\beta=\alpha s$; $T''=T'^{\alpha} H^{r_3}$; $r_5 = r_3 + \alpha r_2 (mod~p)$; $Com=g_1^k h_T^{\nu}$}\\
\mymulticolumn{3}{l}{\textbf{Pick} the valid BB-signature corresponding to the element $k$: $A_k = g^{1/(y+k)}$}\\
\mymulticolumn{2}{l}{\textbf{Choose} $l \in \mathbb{Z}^{*}_p$ and \textbf{Compute:} $B=A_k^{l}$; $B_1=B^{-1}$; $D=B_1^k g^l$}&\\
\mymulticolumn{3}{l}{\textbf{Choose} $k_1, l_1, r_1 \in \mathbb{Z}^{*}_p$ and \textbf{Compute:} $Com_1=g_1^{k_1} h_T^{r_1}$; $D_1=B_1^{k_1} g^{l_1}$}\\
\mymulticolumn{3}{l}{$\delta=s+k$; $f=a+\nu$; $K=g_t B_k^{-1}=B_k^{s+k}=B_k^{\delta}$; $L=C_2Com=g_1^{s+k}h_T^{a+\nu}=g_1^{\delta}h_T^{f}$}\\
&&\\
\mymulticolumn{1}{l}{\underline{\textit{Computation of the witnesses}}} &&\\
\mymulticolumn{3}{l}{\textbf{Compute:} $C'_1=g_T^{a_1}$; $C'_2=g_1^{d_1} \times h_T^{a_1}$; $R'=G^{d_1}H^{t_1}$; $R''=T'^{\alpha_1}H^{t_3}$;}\\
\mymulticolumn{3}{l}{$C'=g_1^{b_1} h^{\alpha_1} B'^{t}$; $T'_4=G^{b_1} H^{t_5}$; $K'=B_k^{d'}$; $L'=g_1^{d'} \times h_T^{f'}$}\\
\hline
\mymulticolumn{3}{c}{\textbf{Real time computations}}\\ \hline
\mymulticolumn{3}{l}{\underline{\textit{Validator authentication}}}\\
\mymulticolumn{1}{l}{\textbf{Choose} $RC_V \in \mathbb{Z}^{*}_p$ and \textbf{Check} that the number}&&\\
\mymulticolumn{1}{l}{ of used m-tickets < $max_{ticket}$} &$\xtwoheadrightarrow{~~~~~~RC_V~~~~~}$& $TS$ = getTimeStamp() \\
&& \textbf{Compute:}\\
\mymulticolumn{1}{l}{\textbf{Verify} $Signature_{RSA}$ and \textbf{Check} that $TS < \mathcal{D}$ }&$\xtwoheadleftarrow{Signature_{RSA}}$&$Signature_{RSA}$, the RSA signature on $RC_V$ and $TS$\\
&&\\
\mymulticolumn{3}{l}{\underline{\textit{m-ticket Validation}}}\\
\mymulticolumn{1}{l}{\textbf{Compute:} $c= H(C, C_1, C_2, B_0, T', T'',Com,B,$}&$\xtwoheadleftarrow{~~~~~~~ch~~~~~~~}$&\textbf{Choose} $ch \in \mathbb{Z}^{*}_p$\\
\mymulticolumn{1}{l}{$D, K, C'_{1}, C'_{2}, R', R'', C', T'_4, Com_1, D_1, K', L', ch)$}&&\\
\mymulticolumn{2}{l}{$s_1=k_1+c\times k~mod~p$; $s_2=r_1+c\times \nu~mod~p$}&\tikz[overlay,remember picture]{\node[anchor=base] (X) {}; }\textbf{Check} that $B \neq 1_{G_1}$\\
\mymulticolumn{2}{l}{$s_3=l_1+c\times l~mod~p$; $\omega_1=a_1+c\times a~mod~p$}&\multirow{3}{\linewidth}{$\bullet$ If The validator  holds the private signature keys $y$ and $\gamma$ \textbf{(First case).} Hence the prover does not send the value $D$ and $C$. The verifier can compute it: $D = B^y$, $C=B_0^{\gamma}$. Then goes to (*)}\\
\mymulticolumn{2}{l}{$\omega_2=d_1+c\times s~mod~p$; $\omega_3=t_1+c\times r_2~mod~p$}&\\
\mymulticolumn{2}{l}{$\omega_4=b_1+c\times \beta~mod~p$; $\omega_5=\alpha_1+c\times \alpha~mod~p$}&\\
\mymulticolumn{2}{l}{$\omega_6=t+c\times r~mod~p$; $\omega_8=t_3+c\times r_3~mod~p$}&\\
\mymulticolumn{2}{l}{$\omega_{10}=t_5+c\times r_5~mod~p$; $\omega_{11}=d'+c \times \delta~mod~p$}&\multirow{2}{\linewidth}{$\bullet$ Otherwise, if the validator doesn't know the private signature key $y$ and $\gamma$ \textbf{(Second case),} then, check that} \\
\mymulticolumn{1}{l}{$\omega_{12}=f'+c\times f~mod~p$} &&\\
\mymulticolumn{1}{l}{Let $E = (C_1, C_2)$ and the proof} &&$e(D, g_3)=e(B, Y)$; $e(C, g_2)=e(B_0, W)$ and goes to (*)\\
\mymulticolumn{1}{l}{$\Pi =(C,B_0,T',T'',Com,c,B,D,s_1,s_2,s_3,$}&&\textbf{(*) Compute:} $\tilde{C}=g_1^{s_1}h_T^{s_2}Com^{-c}$; $\tilde{D}=B_1^{s_1}g^{s_3}D^{-c}$;\\
\mymulticolumn{1}{l}{$\omega_1,\omega_2, \omega_3, \omega_4, \omega_5, \omega_6, \omega_8, \omega_9, \omega_{10}, \omega_{11},\omega_{12})$\tikz[overlay,remember picture]{\node[anchor=base] (Y) {}; 
\draw[->>,color=black,line width=0.25pt] (Y.east) .. controls +(2,0) and +(-2,0) .. node[right]{$B_k,E,\Pi$} (X.north west); }}&&$\tilde{C}_1=g_T^{\omega_1}C_1^{-c}$; $\tilde{C}_2=g_1^{\omega_2}h_T^{\omega_1}C_2^{-c}$; $\tilde{R'}=G^{\omega_2} H^{\omega_3} T'^{-c}$;\\
&&  $\tilde{C'}=g_1^{\omega_4} h^{\omega_5} B'^{\omega_6} C^{-c}$; $\tilde{R''}= T'^{\omega_5} H^{\omega_8} T''^{-c}$;\\
&& $\tilde{T_4}=G^{\omega_4} H^{\omega_10} T''^{-c}$; $\tilde{K'}=B_k^{\omega_{11}} K^{-c}$; $\tilde{L'}=g_1^{\omega_11} h_T^{\omega_12} L^{-c}$\\ 
&&\textbf{Check} $c= H(C, C_1, C_2, B_0, T', T'', Com, B, D, K,$ $\tilde{C_{1}}, \tilde{C_{2}}, \tilde{R'}, \tilde{R''}, \tilde{C'}, \tilde{T'_4}, \tilde{C}, \tilde{D}, \tilde{K'}, \tilde{L'}, ch)$ \\
&&\textbf{Send} $(B_k,E,\Pi, ID_V, DT)$ to TA in order to be saved within the secured database $DB_{UsedTickets}$; $ID_V$ is the identity of the validator and $DT$ is the date and time of the transaction.\\
\end{tabular}
\caption{The validation protocol}
\label{tab:ticketValidation}
\end{figure*}

\textbf{Remark 2:} Proving the knowledge of a valid BB-signature $A =(g_1^s h)^{1/(\gamma+r)}$ on a value $s$, without revealing $s$ or the signature and without using pairings on the prover's side can be done as follows:

The prover first randomizes $A$ by choosing a random value $\alpha \in \mathbb{Z}_p^*$ and computes $B_0=A^{\alpha}$. Since $A =(g_1^s h)^{1/(\gamma+r)}$ this implies that $B_0$ = $A^{\alpha}$ = $(g_1^{\alpha s} h^\alpha)^{1/(\gamma+r)}$ and then that:
\begin{eqnarray}
B_0^{\gamma + r} =  g_1^{\alpha s} h^\alpha
\label{remark1-1} 
\end{eqnarray}
Let us note by $B' = B_0^{-1}$ and $C = B_0^\gamma$.   From~(\ref{remark1-1}), we have:
\begin{eqnarray}
C = g_1^{\alpha s} \times h^{\alpha} \times B'^r 
\label{remark1-2} 
\end{eqnarray}
As a result, in order to prove that he knows a BB-signature $A$ on the value $s$, without revealing $s$ nor the corresponding signature, the prover just has to send $B_0$ and $C$ (that he can compute using~(\ref{remark1-2})) to the verifier and prove that he knows the representation of $C$ with respect to ($g_1$, $h$, $B'$): $\Pi_{BB}$ = $POK$($\alpha$, $s$, $r$ :$C$ = $g_1^{\alpha s}$ $\times$ $h^{\alpha}$ $\times$ $B'^r$). 
The proof consists in $B_0$, $C$ and $\Pi_{BB}$ (and no pairing computations are needed on the prover's side to compute $\Pi_{BB}$). The verifier will have to check that $B_0 \neq 1_{G_1}$, that $C = B_0^\gamma$ (via pairing computations or by using the key $\gamma$ if it owns this key) and that $\Pi_{BB}$  is valid. If all the verifications hold, the verifier will be convinced that the prover knows a valid BB-signature.

As described in Figure~\ref{tab:ticketValidation}, the validator verifies the proof $\Pi$, saves the date and time of the operation $DT$, the serial number of the validated m-ticket $B_{k}$, the El Gamal encryption $E$ (of $g_1^s$) and the proof $\Pi$. 
These verifications can be done in two ways. In the first case, the validator holds the private keys $\gamma$ (the private key of TA) and $y$ (the private key used during the set membership).
Hence, he can perform the verification of $\Pi$ without executing any pairing computations. In such a case, the protocol is run without any pairing computations either on the user side (SIM card) or on the validator side.  In the second case, the validator does \emph{not} hold the private keys $\gamma$ and $y$. Therefore, in order to perform the verification of $\Pi$, the validator would execute pairing computations. We still achieve our goal, i.e., no pairing computations at the user side (SIM card).

We emphasize that owing to our improvements on Boneh-Boyen based Camenisch-Lysyanskaya signature scheme, all the computations (required to validate an m-ticket) are performed by the SIM card. We do not need  to outsource part of the computations to the powerful but untrusted mobile phone. Consequently, the user's privacy is ensured with respect to the smartphone itself.  Even a compromised mobile, e.g. containing a spyware, cannot collect any information about our m-ticketing application computations.

\textbf{Theorem 3.} The protocol in Figure~\ref{tab:ticketValidation} is a ZKPK of a permission token $(A, r, s)$ and a value $k$ such that $B_k=g_t^{1/(s+k+1)}$, $k \in [1.. max_{ticket}]$ and $E=(C_1,C_2)$ is an El Gamal encryption of $g_1^s$. The proof is detailed in Appendix~\ref{appendix:proof3}.

At the end of a successful m-ticket validation, the validator sends $(B_k,E,\Pi)$ to TA in order to be saved within a centralized and secured database $DB_{UsedTickets}$ jointly with his identity $ID_V$ and  the date and time of the transaction $DT$. In such a way, the TA will detect any malicious behaviour such that a multiple usage or cloning (cf. Section~\ref{subsub:couter}).

\subsubsection{Revocation}
We distinguish two levels of revocation: the user's anonymity revocation and the m-tickets revocation. In the first case, the transport authority would like to get the user's identity corresponding to a given m-ticket. In the second case, the transport authority would like to revoke all the m-tickets of a given user, e.g., upon the request of the user further to the theft of his smartphone.    
  
In order to recover a user's identity based on a used m-ticket $Tick_k=(B_k,E,\Pi)$, TA sends $E$ to the revocation authorities. Together, the revocation authorities decipher $E$ and retrieve the commitment of $s$ ($g_1^s$). By using $DB_{REG}$ the database of registered users and $g_1^s$, TA will find the user's identity in the tuples $(A, C_0, s_2, c, \mu, ID_U, h_U)$.

In order to retrieve the m-tickets of a user based on his identity $ID_{U'}$, first of all, TA must find the tuple $(A, C_0, s_2, c,$ $\mu, ID_U, h_U)$ from $DB_{REG}$ such that $ID_U = ID_{U'}$. Then, TA sends $C_0$ (the Paillier encryption of $s_1$) to the revocation authorities. Similarly to the first case, the revocation authorities decipher together $C$ and retrieve $s_1$. Upon receiving $s_1$, TA computes $s=s_1+s_2$, hence, it can retrieve all the m-tickets $(B_k=g_t^{1/(s+k+1)})$ of the user.

\subsection{A secure Post-payment process}
\label{sec:PPP}
One novelty of our m-ticketing protocol, in addition to the respect of users' privacy, is to give the ability to charge a user after the usage of m-tickets.

\begin{table*}
\scriptsize
\begin{center}
\begin{tabular}{ccccccc}
& Validator & Card Signature  & \multicolumn{2}{c}{Verification by PC} & \multicolumn{2}{c}{Total} \\ 
& authentication  & + NFC connection & (1) without pairing& (2) with pairing& (1) & (2)\tabularnewline

\hline
\multicolumn{1}{c}{Battery-On}  &$56,98(0.70)$&$123.01(3.24)$  & \multirow{2}{*}{$4.43(1.32)$} & \multirow{2}{*}{$12.19(3.20)$} & $184.25(3.43)$ &  $191.80(4.73)$\\
\multicolumn{1}{c}{Battery-Off}& $76.55(7.46)$ &$185.28(18.68)$  &       &      & $266.52(17.91)$ & $272.55(25.73)$ \\
\end{tabular}
\end{center}
\caption{Timings of m-ticket validation protocol including validator authentication (ms)}
\label{tab:timing}
\end{table*}

\subsubsection{Regular reporting}
In a post-payment approach, the m-ticketing application must report unused m-tickets to the back-end server, before the pre-defined $D$ day. Thus, the regular reports does not question the privacy of the user. The following example gives further clarification. Suppose that the user retrieved a permission token of $5$ m-tickets. Before $D$ day, he used 4 m-tickets, i.e., m-tickets number 1, 2, 3 and 4.  On $D$ day, the m-ticketing application will report to the back-end server the m-ticket number 5 using a network connection. The report contains $B_{5}$ = $g_{t}^{1/s+5+1}$ and  the proof $\Pi= POK(k,s,r,A: B_5=g_t^{1/(s+5+1)} \wedge A=(g_1^s h)^{1/(\gamma+r)} \wedge 5 \in [1..5])$. 

Regularly revealing the unused m-tickets enables the transport authority to (1) charge the user, (2) check the reliability and accuracy of the reports without questioning the user's privacy. Indeed, the unlinkability of user's m-tickets $(B_k=g_t^{1/(s+k+1)}, E, \Pi)$, both during the use of the service and during the reporting phase, is ensured owing to the q-DDHI and DDH assumptions (cf. \textit{unlinkabililty} proof).

\subsubsection{Countermeasures}
\label{subsub:couter}
A malicious user's goal simply consists on not paying or paying less than what he is supposed to. To do so, he may try to block the reporting or attack the cardlet of the SIM.

If the user blocks network communications in order to block reporting, the SIM cardlet will refuse to issue m-tickets for a permission token after reaching the limit of $max_{ticket}$, or the limit of the time window controlled by $\mathcal{D}$. This countermeasure $C$ relies on the security of the SIM card.

If the user performs a successful physical attack against the SIM card, which is extremely difficult~\cite{SAT10}, he may try to (1) defeat the countermeasure $C$ and never do reports, or (2) use the same m-ticket several times or (3) report a used m-ticket. This will be detected in $DB_{UsedTickets}$ and TA can decide to break the user's anonymity and retrieve all his m-tickets usage history.

Thus, the user is forced to report his consumption and renew his permission token lest the service is unusable or TA breaks his anonymity. 

Note that forging an m-ticket is not possible, even with a compromised SIM because of the \textit{unforgeability} property.

\section{Security analysis}
\label{sec:analysis}
We prove that our m-ticketing protocol provides the security and privacy properties defined in Section~\ref{sub:secmodel}.

\textbf{Theorem 4 (Non-frameability).} Our m-ticketing protocol is non-frameable, in the random oracle model, under the q-DDHI assumption. 

The proof is detailed is Appendix~\ref{appendix:proofs}.

\textbf{Theorem 5 (Unforgeability).} Our m-ticketing protocol satisfies the unforgeability requirement, in the random oracle model, under the q-SDH assumption.

The proof is detailed is Appendix~\ref{appendix:proofs}.

\textbf{Theorem 6 (Unlinkability).} Our m-ticketing protocol satisfies the unlinkability requirement, in the random oracle model, under the q-DDHI and DDH assumptions.

\begin{fullversion}
The proof is detailed is Appendix~\ref{appendix:proofs}.
\end{fullversion}
\begin{petsversion}
Because of space limitation, the proof of Theorem~6  is omitted from this paper but is detailed in the extended version~\cite{ALT+15}. 
\end{petsversion}

\section{Performance results}
\label{sec:PoC}
We implemented the user side of our solution on a SIM card and the validator side in a regular PC (cf. Appendix~\ref{appendix:param}). 
We used a 256-bit Barreto-Naehrig curve~\cite{BPN06} over $\mathbb{F}_{q}$ since this family of curves provides an optimal size for $\mathbb{G}_{1}$ and $\mathbb{G}_{2}$ while preventing the MOV attack~\cite{MVO91} due to their embedding degree of 12. $G_{1}$ is the group of $\mathbb{F}_{q}$ - rational points
of order p and $G_{2}$ is the subgroup of trace zero points in $E(F_{q^{12}})[p]$. Our pairing is of type-3~\cite{SKN08}. More details are given in Appendix~\ref{appendix:param}.

\textbf{Pre-computations.} 
The pre-computations for preparing one signature occurs in $1.7\,s$. It consists in computing elements involved in proving that the user knows a valid signature on his secret $s$. The total size of the elements that the card has to store for one signature is $1130~bytes$ (24 $\mathbb{Z}_{p}$ elements, $10$ compressed points and one digest output). 

\textbf{Validation phase.}
Table~\ref{tab:timing} gives timings (average over 50 trials and standard deviation between parentheses) for the whole validation protocol which includes the validator authentication, the signature generation and an m-ticket verification. The timings include as well the NFC exchanges duration. We denote by ``Battery-Off'' a powered-off phone either by the user, or because the battery is flat. In this situation, as stated by NFC standards, NFC-access to the SIM card is still possible, but with degraded performances. 

Regarding the validator authentication, we chose to use RSA with a $1984$ bits key (this is the greatest size supported by the SIM card) and a short public verification exponent $v$ ($v =65537$). The validator asks the card for a challenge ($RC_V$). Then, he sends back his response to this challenge, his own challenge ($ch$) ($32$ bytes) and the current date ($TS$) ($6$ bytes). 

The column ``Card signature'' gives the duration of real-time computations of a signature (computing an m-ticket $B_k$, $E$ and $\Pi$) and the NFC communication time. The considered operations for generating the signature  are only one hash value and lightweight operations in $\mathbb{Z}_{p}$. The size of the computed signature is $778~bytes$ (sent by 4 APDUs). Regarding the communication between a SIM card and a reader, it is slow ($\ge 85\,ms$), but the whole process (Signature+NFC) remains very fast, i.e., $123.01\,ms$ on average.

For the signature verification by the validator, we distinguish two cases: the validator holds the private signature keys ($y$, $\gamma$), or not. The extra pairing computations performed in the second case do not add an important delay to the verification because a regular desktop PC can efficiently achieve such computations.

In total, the m-ticket validation occurs for the first case (without pairing at the verifier side) in $184.25\,ms$ on average and in $191.80\,ms$ for the second case (with pairing at the verifier side). When the battery is flat, the validation occurs in at most $272.55\,ms$ on average.

\section{Conclusion}
\label{sec:conclusion}
In this paper, we proposed various cryptographic enhancements: (1) optimizations of BB-signature schemes and (2) a new set-membership proof that does not require pairing computations at the prover side. These contributions enable to design efficient protocols over NFC. Then, based on these cryptographic primitives, we designed a secure m-ticketing protocol that prevents a malicious transport authority from linking users' trips. Moreover, as the entire computations are done within the SIM card, we enhance the user's privacy with regards to a compromised smartphone. Owing to regular reports of unused m-tickets, the user is able to securely post-pay for his m-tickets without disclosing any information about the performed trips. Moreover, we planned countermeasures against m-ticket cloning and multiple usage. Finally, we prove that our protocol is completely off-line and efficient: the validation occurs in $184.25\,ms$ and in $266.52\,ms$ when the battery is flat.



\begin{thebibliography}{10}


\bibitem{BPN06}
P.~S. Barreto and M.~Naehrig.
\newblock {Pairing-Friendly Elliptic Curves of Prime Order}.
\newblock In B.~Preneel and S.~Tavares, editors, {\em Selected Areas in
  Cryptography}, volume 3897 of {\em LNCS}, pages 319--331. Springer Berlin
  Heidelberg, Kingston, ON, Canada, 2006.

\bibitem{BPS03}
Bellare, Namprempre, Pointcheval, and Semanko.
\newblock The one-more-{RSA}-inversion problems and the security of chaum's
  blind signature scheme.
\newblock {\em Journal of Cryptology}, 16(3):185--215, 2003.
\newblock DOI:
  \href{http://dx.doi.org/10.1007/s00145-002-0120-1}{10.1007/s00145-002-0120-1}.

\bibitem{BR93}
M.~Bellare and P.~Rogaway.
\newblock {Random Oracles Are Practical: A Paradigm for Designing Efficient
  Protocols}.
\newblock In {\em 1st ACM Conference on Computer and Communications Security},
  CCS '93, pages 62--73, Fairfax, Virginia, USA, 1993. ACM.

\bibitem{berlin}
Berlin.de.
\newblock Tickets, fares and route maps.
\newblock \url
  {http://www.berlin.de/en/public-transportation/1772016-2913840-tickets-fares-and-route-maps.en.html}.

\bibitem{EAR13}
E.-O. Blass, A.~Kurmus, R.~Molva, and T.~Strufe.
\newblock {PSP: Private and secure payment with {RFID}}.
\newblock {\em Computer Communications}, 36(4):468--480, 2013.
\newblock DOI:
  \href{http://dx.doi.org/10.1016/j.comcom.2012.10.012}{10.1016/j.comcom.2012.10.012}.

\bibitem{BB04}
D.~Boneh and X.~Boyen.
\newblock {Short Signatures Without Random Oracles}.
\newblock In C.~Cachin and J.~L. Camenisch, editors, {\em Advances in
  Cryptology - EUROCRYPT 2004}, volume 3027 of {\em LNCS}, pages 56--73.
  Springer Berlin Heidelberg, Interlaken, Switzerland, 2004.

\bibitem{BB08}
D.~Boneh and X.~Boyen.
\newblock {Short Signatures Without Random Oracles and the SDH Assumption in
  Bilinear Groups}.
\newblock {\em Journal of Cryptology}, 21(2):149--177, 2008.
\newblock DOI:
  \href{http://dx.doi.org/10.1007/s00145-007-9005-7}{10.1007/s00145-007-9005-7}.

\bibitem{BBS04}
D.~Boneh, X.~Boyen, and H.~Shacham.
\newblock {Short Group Signatures}.
\newblock In M.~Franklin, editor, {\em Advances in Cryptology - CRYPTO '04},
  volume 3152 of {\em LNCS}, pages 41--55. Springer Berlin Heidelberg, Santa
  Barbara, California, USA, 2004.

\bibitem{CCS08}
J.~Camenisch, R.~Chaabouni, and A.~Shelat.
\newblock {Efficient Protocols for Set Membership and Range Proofs}.
\newblock In J.~Pieprzyk, editor, {\em Advances in Cryptology - ASIACRYPT
  2008}, volume 5350 of {\em LNCS}, pages 234--252. Springer Berlin Heidelberg,
  Melbourne, Australia, 2008.

\bibitem{CL04}
J.~Camenisch and A.~Lysyanskaya.
\newblock {Signature Schemes and Anonymous Credentials from Bilinear Maps}.
\newblock In M.~Franklin, editor, {\em Advances in Cryptology - CRYPTO '04},
  volume 3152 of {\em LNCS}, pages 56--72. Springer Berlin Heidelberg, Santa
  Barbara, California, USA, 2004.

\bibitem{CPS96}
J.~Camenisch, J.-M. Piveteau, and M.~Stadler.
\newblock {An Efficient Fair Payment System}.
\newblock In {\em 3rd ACM Conference on Computer and Communications Security},
  CCS '96, pages 88--94, New Delhi, India, 1996. ACM.

\bibitem{CCJ13}
S.~Canard, I.~Coisel, A.~Jambert, and J.~Traor\'e.
\newblock {New Results for the Practical Use of Range Proofs}.
\newblock In S.~Katsikas and I.~Agudo, editors, {\em Public Key
  Infrastructures, Services and Applications}, volume 8341 of {\em LNCS}, pages
  47--64. Springer Berlin Heidelberg, Egham, UK, 2014.

\bibitem{CLZ12}
R.~Chaabouni, H.~Lipmaa, and B.~Zhang.
\newblock {A Non-interactive Range Proof with Constant Communication}.
\newblock In A.~Keromytis, editor, {\em Financial Cryptography and Data
  Security}, volume 7397 of {\em LNCS}, pages 179--199. Springer Berlin
  Heidelberg, Kralendijk, Bonaire, 2012.

\bibitem{CP92}
D.~Chaum and T.~P. Pedersen.
\newblock {Wallet Databases with Observers}.
\newblock In E.~F. Brickell, editor, {\em Advances in Cryptology - CRYPTO '92},
  volume 740 of {\em LNCS}, pages 89--105, Santa Barbara, California, USA,
  1993. Springer Berlin Heidelberg.

\bibitem{Chaumette2011}
S.~Chaumette, D.~Dubernet, and J.~Ouoba.
\newblock {Architecture and comparison of two different user-centric
  NFC-enabled event ticketing approaches}.
\newblock In S.~Balandin, Y.~Koucheryavy, and H.~Hu, editors, {\em The 11th
  international conference on next generation wired/wireless networking},
  volume 6869 of {\em LNCS}, pages 165--177, St. Petersburg, Russia, 2011.
  Springer Berlin Heidelberg.

\bibitem{DKJ12}
D.~Derler, K.~Potzmader, J.~Winter, and K.~Dietrich.
\newblock {Anonymous Ticketing for NFC-Enabled Mobile Phones}.
\newblock In L.~Chen, M.~Yung, and L.~Zhu, editors, {\em Trusted Systems},
  volume 7222 of {\em LNCS}, pages 66--83, Beijing, China, 2012. Springer
  Berlin Heidelberg.

\bibitem{DF89}
Y.~G. Desmedt and Y.~Frankel.
\newblock Threshold cryptosystems.
\newblock In G.~Brassard, editor, {\em Advances in Cryptology - CRYPTO '89},
  volume 435 of {\em LNCS}, pages 307--315, Santa Barbara, California, USA,
  1989. Springer Berlin Heidelberg.

\bibitem{DSD07}
A.~Devegili, M.~Scott, and R.~Dahab.
\newblock {Implementing Cryptographic Pairings over Barreto-Naehrig Curves}.
\newblock In T.~Takagi, T.~Okamoto, E.~Okamoto, and T.~Okamoto, editors, {\em
  Pairing-Based Cryptography - Pairing 2007}, volume 4575 of {\em LNCS}, pages
  197--207. Springer Berlin Heidelberg, Tokyo, Japan, July 2007.

\bibitem{DST12}
A.~Dmitrienko, A.-R. Sadeghi, S.~Tamrakar, and C.~Wachsmann.
\newblock {SmartTokens: Delegable Access Control with {NFC}-Enabled
  Smartphones}.
\newblock In S.~Katzenbeisser, E.~Weippl, L.~Camp, M.~Volkamer, M.~Reiter, and
  X.~Zhang, editors, {\em Trust and Trustworthy Computing}, volume 7344 of {\em
  LNCS}, pages 219--238. Springer Berlin Heidelberg, Vienna, Austria, 2012.

\bibitem{DY03}
Y.~Dodis.
\newblock {Efficient Construction of (Distributed) Verifiable Random
  Functions}.
\newblock In Y.~Desmedt, editor, {\em Public Key Cryptography - PKC 2003},
  volume 2567 of {\em LNCS}, pages 1--17. Springer Berlin Heidelberg, Miami,
  FL, USA, 2003.

\bibitem{DY05}
Y.~Dodis and A.~Yampolskiy.
\newblock {A Verifiable Random Function with Short Proofs and Keys}.
\newblock In S.~Vaudenay, editor, {\em Public Key Cryptography - PKC 2005},
  volume 3386 of {\em LNCS}, pages 416--431. Springer Berlin Heidelberg,
  Diablerets, Switzerland, 2005.

\bibitem{Ekberg2012}
J.-E. Ekberg and S.~Tamrakar.
\newblock {Mass Transit Ticketing with NFC Mobile Phones}.
\newblock In L.~Chen, M.~Yung, and L.~Zhu, editors, {\em Third International
  Conference on Trusted Systems}, volume 7222 of {\em LNCS}, pages 48--65,
  Beijing, China, 2012. Springer Berlin Heidelberg.

\bibitem{ElGamal85}
T.~El~Gamal.
\newblock {A Public Key Cryptosystem and a Signature Scheme Based on Discrete
  Logarithms}.
\newblock In G.~R. Blakley and D.~Chaum, editors, {\em Advances in Cryptology -
  CRYPTO '84}, volume 196 of {\em LNCS}, pages 10--18, Santa Barbara,
  California, USA, 1985. Springer Berlin Heidelberg.

\bibitem{SAT10}
M.~Eznack, J.-P. Warry, C.~Loiseaux, G.~Dufay, R.~Atoui, N.~Herbreteau,
  J.~Pieniazek, and F.~Thabaret.
\newblock {(U)SIM Java Card Platform Protection Profile Basic and SCWS
  Configurations-Evolutive Certification Scheme for (U)SIM cards, Version
  2.0.2}.
\newblock \url
  {http://www.ssi.gouv.fr/IMG/certificat/ANSSI-CC-cible_PP-2010-04en.pdf}, June
  2010.

\bibitem{FS87}
A.~Fiat and A.~Shamir.
\newblock {How to Prove Yourself: Practical Solutions to Identification and
  Signature Problems}.
\newblock In A.~M. Odlyzko, editor, {\em Advances in Cryptology - CRYPTO '86},
  volume 263 of {\em LNCS}, pages 186--194, Santa Barbara, California, USA,
  1987. Springer Berlin Heidelberg.

\bibitem{PJ01}
P.~Fouque and J.~Stern.
\newblock Fully distributed threshold {RSA} under standard assumptions.
\newblock In C.~Boyd, editor, {\em Advances in Cryptology - ASIACRYPT 2001},
  volume 2248 of {\em LNCS}, pages 310--330, Gold Coast, Australia, 2001.
  Springer Berlin Heidelberg.

\bibitem{nfcSTAT}
{FROST \& SULLIVAN}.
\newblock {NFC: When will be the real start?}
\newblock
  \url{http://www.frost.com/sublib/display-report.do?id=9843-00-13-00-00},
  January 2011.

\bibitem{FPV09}
G.~Fuchsbauer, D.~Pointcheval, and D.~Vergnaud.
\newblock {Transferable Constant-Size Fair E-Cash}.
\newblock In J.~Garay, A.~Miyaji, and A.~Otsuka, editors, {\em Cryptology and
  Network Security}, volume 5888 of {\em LNCS}, pages 226--247. Springer Berlin
  Heidelberg, Kanazawa, Japan, 2009.

\bibitem{SKN08}
S.~D. Galbraith, K.~G. Paterson, and N.~P. Smart.
\newblock {Pairings for cryptographers }.
\newblock {\em Discrete Applied Mathematics}, 156(16):3113--3121, 2008.
\newblock DOI:
  \href{http://dx.doi.org/10.1016/j.dam.2007.12.010}{10.1016/j.dam.2007.12.010}.

\bibitem{GJK03}
R.~Gennaro, S.~Jarecki, H.~Krawczyk, and T.~Rabin.
\newblock {Secure Applications of Pedersen Distributed Key Generation
  Protocol}.
\newblock In M.~Joye, editor, {\em Topics in Cryptology - CT-RSA 2003}, volume
  2612 of {\em LNCS}, pages 373--390. Springer Berlin Heidelberg, San
  Francisco, CA, USA, 2003.

\bibitem{CLJ12}
C.~P.~L. Gouv{\^{e}}a, L.~B. Oliveira, and J.~L{\'{o}}pez.
\newblock {Efficient software implementation of public-key cryptography on
  sensor networks using the {MSP430X} microcontroller}.
\newblock {\em J. Cryptographic Engineering}, 2(1):19--29, 2012.
\newblock DOI:
  \href{http://dx.doi.org/10.1007/s13389-012-0029-z}{10.1007/s13389-012-0029-z}.

\bibitem{gsma}
{GSMA Mobile NFC}.
\newblock {White Paper: Mobile NFC in Transport}.
\newblock \url
  {http://www.uitp.org/public-transport/technology/Mobile-NFC-in-Transport.pdf},
  September 2012.

\bibitem{HCD06}
T.~S. Heydt-Benjamin, H.-J. Chae, B.~Defend, and K.~Fu.
\newblock {Privacy for Public Transportation}.
\newblock In G.~Danezis and P.~Golle, editors, {\em 6th International
  Conference on Privacy Enhancing Technologies - PET'06}, volume 4258 of {\em
  LNCS}, pages 1--19, Cambridge, UK, 2006. Springer Berlin Heidelberg.

\bibitem{EJ07}
E.~Hufschmitt and J.~Traor{\'{e}}.
\newblock {Fair Blind Signatures Revisited}.
\newblock In T.~Takagi, T.~Okamoto, E.~Okamoto, and T.~Okamoto, editors, {\em
  Pairing-Based Cryptography - Pairing 2007}, volume 4575 of {\em LNCS}, pages
  268--292, Tokyo, Japan, July 2007.

\bibitem{IVM12}
A.~P. Isern-Deya, A.~Vives-Guasch, M.~Mut-Puigserver, M.~Payeras-Capella, and
  J.~Castella-Roca.
\newblock {A Secure Automatic Fare Collection System for Time-Based or
  Distance-Based Services with Revocable Anonymity for Users}.
\newblock {\em The Computer Journal}, 56(10):1198--1215, Apr. 2012.
\newblock DOI:
  \href{http://dx.doi.org/10.1093/comjnl/bxs033}{10.1093/comjnl/bxs033}.

\bibitem{iso14443}
{ISO 14443-3:2011}.
\newblock {Identification cards -- Contactless integrated circuit cards --
  Proximity cards}.

\bibitem{MVO91}
A.~Menezes, S.~Vanstone, and T.~Okamoto.
\newblock {Reducing Elliptic Curve Logarithms to Logarithms in a Finite Field}.
\newblock In {\em 23rd Annual ACM Symposium on Theory of Computing - STOC '91},
  pages 80--89, New Orleans, Louisiana, USA, 1991. ACM.

\bibitem{moscow}
Moscow.
\newblock \url
  {http://moscow.ru/fr/guide/trip_planning/inner_transport/transport/metro/}.

\bibitem{nfc_public_transport}
{NFC Forum}.
\newblock {NFC in Public Transport}.
\newblock \url
  {http://nfc-forum.org/wp-content/uploads/2013/12/NFC-in-Public-Transport.pdf},
  2011.

\bibitem{PAI99}
P.~Paillier.
\newblock {Low-Cost Double-Size Modular Exponentiation or How to Stretch Your
  Cryptoprocessor}.
\newblock In {\em Second International Workshop on Practice and Theory in
  Public Key Cryptography, {PKC} '99}, volume 1560 of {\em LNCS}, pages
  223--234, Kamakura, Japan, Mar. 1999. Springer Berlin Heidelberg.

\bibitem{Pedersen92}
T.~Pedersen.
\newblock { }.
\newblock In J.~Feigenbaum, editor, {\em Advances in Cryptology - CRYPTO '91},
  volume 576 of {\em LNCS}, pages 129--140. Springer Berlin Heidelberg, 1992.

\bibitem{PS96}
D.~Pointcheval and J.~Stern.
\newblock {Security Proofs for Signature Schemes}.
\newblock In U.~Maurer, editor, {\em Advances in Cryptology - EUROCRYPT '96},
  volume 1070 of {\em LNCS}, pages 387--398. Springer Berlin Heidelberg,
  Saragossa, Spain, 1996.

\bibitem{PS00}
D.~Pointcheval and J.~Stern.
\newblock Security arguments for digital signatures and blind signatures.
\newblock {\em J. Cryptology}, 13(3):361--396, 2000.
\newblock DOI:
  \href{http://dx.doi.org/10.1007/s001450010003}{10.1007/s001450010003}.

\bibitem{AFG13}
A.~Rupp, G.~Hinterw\"alder, F.~Baldimtsi, and C.~Paar.
\newblock {P4R: Privacy-Preserving Pre-Payments with Refunds for Transportation
  Systems}.
\newblock In A.-R. Sadeghi, editor, {\em Financial Cryptography and Data
  Security}, volume 7859 of {\em LNCS}, pages 205--212. Springer Berlin
  Heidelberg, Okinawa, Japan, 2013.

\bibitem{AIC08}
A.~Sadeghi, I.~Visconti, and C.~Wachsmann.
\newblock {User Privacy in Transport Systems Based on RFID E-Tickets}.
\newblock In C.~Bettini, S.~Jajodia, P.~Samarati, and X.~S. Wang, editors, {\em
  International Workshop on Privacy in Location-Based Applications - PilBA
  2008}, volume 397, Malaga, Spain, Oct. 2008. CEUR.

\bibitem{CPS91}
C.~Schnorr.
\newblock {Efficient Signature Generation by Smart Cards}.
\newblock {\em Journal of Cryptology}, 4(3):161--174, 1991.
\newblock DOI: \href{http://dx.doi.org/10.1007/BF00196725}{10.1007/BF00196725}.

\bibitem{mob-nfc}
{Smart Card Alliance}.
\newblock Proximity mobile payments: Leveraging {NFC} and the contactless
  financial payments infrastructure.
\newblock \url
  {http://www.smartcardalliance.org/resources/lib/Proximity_Mobile_Payments_200709.pdf},
  2007.

\bibitem{SOS08}
P.~Szczechowiak, L.~Oliveira, M.~Scott, M.~Collier, and R.~Dahab.
\newblock {NanoECC: Testing the Limits of Elliptic Curve Cryptography in Sensor
  Networks}.
\newblock In R.~Verdone, editor, {\em Wireless Sensor Networks}, volume 4913 of
  {\em LNCS}, pages 305--320. Springer Berlin Heidelberg, Bologna, Italy, 2008.

\bibitem{Tamrakar2013}
S.~Tamrakar and J.-E. Ekberg.
\newblock {Tapping and Tripping with NFC}.
\newblock In {\em 6th International Conference on Trust \& Trustworthy
  Computing}, volume 7904 of {\em LNCS}, pages 115--132, London, United
  Kingdom, 2013. Springer Berlin Heidelberg.

\bibitem{paris}
{The Paris Convention and Visitors Bureau}.
\newblock Public transport in paris.
\newblock
  \url{http://en.parisinfo.com/practical-paris/how-to-get-to-and-around-paris/public-transport-paris}.


\begin{petsversion}
\bibitem{ALT+15}
G.~Arfaoui, J.-F. Lalande, J.~Traor\'e, N.~Desmoulins, P.~Berthom\'e, and
  S.~Gharout.
\newblock {arXiv: \href{http://arxiv.org/abs/XXXX.YYYY}{XXXX.YYYY}}.
\end{petsversion}

\begin{fullversion}
\bibitem{BLS04}
D.~Boneh, B.~Lynn, and H.~Shacham.
\newblock Short signatures from the weil pairing.
\newblock {\em Journal of Cryptology}, 17(4):297--319, 2004.
\newblock DOI:
  \href{http://dx.doi.org/10.1007/s00145-004-0314-9}{10.1007/s00145-004-0314-9}.

\bibitem{brands}
S.~Brands.
\newblock {Rethinking Public Key Infrastructures and Digital Certificates:
  Building in Privacy}.
\newblock {\em MIT Press}, 2000.

\bibitem{FP01}
P.-A. Fouque and D.~Pointcheval.
\newblock {Threshold Cryptosystems Secure against Chosen-Ciphertext Attacks}.
\newblock In C.~Boyd, editor, {\em Advances in Cryptology - ASIACRYPT 2001},
  volume 2248 of {\em LNCS}, pages 351--368. Springer Berlin Heidelberg, Gold
  Coast, Australia, 2001.

\bibitem{MG13}
M.~Geuss.
\newblock {Japanese railway company plans to sell data from e-ticket records.
  Ars Technica}.
\newblock \url{http://ars.to/13t01BC}, July 2013.

\bibitem{Shoup04}
V.~Shoup.
\newblock {Sequences of Games: A Tool for Taming Complexity in Security
  Proofs}.
\newblock {\em {IACR} Cryptology ePrint Archive}, 2004, 2004.

\end{fullversion}
%
\end{thebibliography}


\appendix

\begin{fullversion}

\section{Privacy weakness of Rupp et al. protocol}
\label{appendix:flaw}
In the following, we give further details about the payment protocol for public transport proposed by Rupp et al. at FC 2013~\cite{AFG13}. Then, we present an attack that enables a malicious transport authority to break users' privacy.

At first, the user buys a trip authorization token ($TAT_i$) which consists of an (extended) coin in Brands' scheme~\cite{brands}. This token enables the user to perform one trip. Then, the user receives a fresh refund token ($RT=SN_{RT}$~such that $SN_{RT}\leftarrow G$), sets a variable $R$ to 1 and  $\upsilon$ to 0. $R$ will be used to aggregate blinding factors and $\upsilon$ is the refund amount. At the entry turnstile, the user shows a $TAT_i$ and proves its ownership. Right after this, the user receives a refund calculation token ($RCT_i$) which consists of a MAC on $TAT_i$, a timestamp and the reader ID. At the turnstile of the exit, the user shows his $RCT$ and proves the ownership of the $TAT_i$ contained within it. In addition, he collects refund on his $RT$, as described in Figure~\ref{tab:refundCollect}. A refund value $\omega$ is represented as a $\omega$-times BLS signature~\cite{BLS04} $SN_{RT}^{d^\omega}$ such that $d$ is the secret BLS signature key of the transport authority. Later, the user redeems his $RT$. This consists on sending a blinded version of his $RT$ ($RT'=RT^r$), $SN_{RT}$, $R$ and the collected amount $\upsilon$ to the transport authority. The latter checks the validity of both $SN_{RT}$ and the signature:
\begin{eqnarray}
e(SN_{RT}^{R}, h^{d^\upsilon})\overset{?}{=}e(RT',h)
\label{eq1}
\end{eqnarray}
In~\cite{AFG13}, Rupp et al. assume that, for efficiency reasons, users don't verify their RTs. Indeed, verifications of refund tokens (step 1 in Figure~\ref{tab:refundCollect}) is too costly and can not be handled by constrained devices such as SIM cards. This lack of verification could lead to serious privacy weaknesses: a malicious transport authority could link users' trips.

\begin{figure}
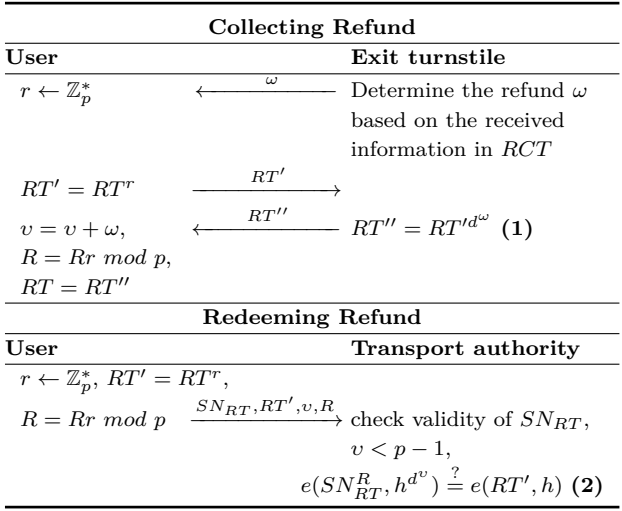

\scriptsize
\centering
\begin{tabular}{>{\normalfont\mathversion{normal}\raggedright}p{\dimexpr 0.3\linewidth-\tabcolsep}
>{\normalfont\mathversion{normal}\raggedleft}p{\dimexpr 0.2\linewidth-\tabcolsep}
>{\normalfont\mathversion{normal}\raggedright\arraybackslash}p{\dimexpr 0.45\linewidth-\tabcolsep}}
\mymulticolumn{3}{c}{\textbf{Collecting Refund}\rule[-3pt]{0pt}{12pt}}\\
\hline
\textbf{User}&&\textbf{Exit turnstile}\\
\hline
\mymulticolumn{1}{l}{$r\leftarrow\mathbb{Z}_p^*$}&\makebox[1.5cm]{$\xleftarrow{~~~~~~~~\omega~~~~~~~}$}&Determine the refund $\omega$ based on the received information in $RCT$ \\
\mymulticolumn{1}{l}{$RT' = RT^r$}&\makebox[1.5cm]{$\xrightarrow{~~~~~~~RT'~~~~~~}$}&\\
\mymulticolumn{1}{l}{$\upsilon=\upsilon+\omega$,}&\makebox[1.5cm]{$\xleftarrow{~~~~~~RT''~~~~~~}$}&$RT''=RT'^{d^\omega}$ \textbf{(1)}\\
\mymulticolumn{1}{l}{$R=Rr~mod~p$,}&&\\
\mymulticolumn{1}{l}{$RT=RT''$}&&\\
\hline
\mymulticolumn{3}{c}{\textbf{Redeeming Refund}}\\
\hline
\textbf{User}&&\textbf{Transport authority}\\
\hline
\mymulticolumn{2}{l}{$r\leftarrow\mathbb{Z}_p^*$, $RT'=RT^r$,}&\\
\mymulticolumn{1}{l}{$R=Rr~mod~p$}&\makebox[1.5cm]{$\xrightarrow{SN_{RT}, RT', \upsilon, R}$}& check validity of $SN_{RT}$,\\
&& $\upsilon<p-1$,\\
&\mymulticolumn{2}{r}{$e(SN_{RT}^{R}, h^{d^\upsilon})\overset{?}{=}e(RT',h)$ \textbf{(2)}}\\
\end{tabular} 
\caption{The refund protocols of Rupp et al.~\cite{AFG13}}
\label{tab:refundCollect}
\end{figure}

More precisely, for a given station, the transport authority is able to identify all the users who left at this station. To this end, in step (1) in Figure~\ref{tab:refundCollect}, instead of computing $RT''=RT'^{d^\omega}$, the transport authority chooses a variable $t\leftarrow\mathbb{Z}_p^*$ and computes  $RT''$=$RT'^{\textbf{t}d^\omega}$. The user will not detect that $RT''$ is improperly calculated because there is no verification on the user's side. The refund token $RT$ will then carry on the exponent $t$ till the redeeming refund step, i.e., $RT' = SN_{RT}^{\textbf{t}^{\textbf{j}} R d^{\upsilon}}$ where $j$ corresponds to the number of times the user left the targeted station, $R$ is the product of all the variables $r$ and $\upsilon$ is the sum of all the refunds $\omega$. At the redeeming refund step, upon receiving the serial number $SN_{RT}$, the blinded version of the refund token $RT'$, the refund amount $\upsilon$, and the aggregate blinding factor $R$, the transport authority will as usual check the validity of $SN_{RT}$ and whether the amount is within the allowed range $[0, p-1]$. Now, in order to verify the signature and distinguish users who exited at the relevant station, the transport authority starts by checking the relation~1\ref{eq1}). If equation~(\ref{eq1}) holds, this implies that the user did not left the targeted station. Otherwise, the transport authority checks: 
\begin{eqnarray}
e(SN_{RT}^{\textbf{t}^{\textbf{j}}R}, h^{d^\upsilon})= {e(SN_{RT}, h)}^{\textbf{t}^{\textbf{j}}Rd^\upsilon}= \nonumber \\ e(SN_{RT}^{\textbf{t}^{\textbf{j}} R d^{\upsilon}}, h)=e(RT',h)
\label{eq2} 
\end{eqnarray}
To do so, the transport authority will repeat the computations with different values of $j$ until the relation~(\ref{eq2}) holds. When $e(SN_{RT}^{t^jR}, h^{d^\upsilon}) \overset{?}{=} e(RT',h)$ equation~(\ref{eq2}) holds, this implies that the user left the targeted station $j$ times. 

In order to monitor several stations at the same time, the transport authority will have a list of variables $t$, e.g., $t_A$ for station A, $t_B$ for station B, etc. At the redeeming refund step, the transport authority will test different values of $t$ until it finds the ones that satisfy the relation~(\ref{eq3}):
\begin{eqnarray}
e(SN_{RT}^{\prod t_{x}^{j_x}R}, h^{d^\upsilon}) \overset{?}{=} e(RT',h)
\label{eq3} 
\end{eqnarray}
where $t_{x}$ characterizes a station $x$ and $j_x$ represents the number of exits at the station $x$. For instance, if the user left the station A four times, the station B twice and the station M once, relation~(\ref{eq3}) will be as follows: $e(SN_{RT}^{ t_{A}^{4}t_{B}^{2}t_{M}^{1}R}, h^{d^\upsilon}) \overset{?}{=} e(RT',h)$. In this way, the transport authority will be able to identify the users that left a pool of targeted stations but also the number of times they exited at these stations: this clearly implies that users' trips are \emph{linkable}.

However, solving relation~(\ref{eq3}) and finding the right tuples $(t_x, j_x)$, becomes quickly a complex and unmanageable task when the number of targeted stations increases (more than three for example). To mitigate this, a malicious transport authority could only monitor two stations at the same time and periodically change these monitored stations. In this way, the transport authority will be able to monitor a large number of stations and will have a global overview of the users' journeys which breaks the privacy property of Rupp et al. system. We would like to emphasize that privacy holds in P4R (Rupp et al. system~\cite{AFG13}) if we assume that the transport authority is ``honest but curious'', i.e., it will not deviate from the refund protocol, but such an assumption is too strong in the context of transport systems~\cite{MG13}.

In order to mitigate this issue, the user should check the received refund (Step (1) in Figure~\ref{tab:refundCollect}). This verification implies pairing computations. However, in constraint environments such as the SIM card, this is not feasible yet. A straightforward solution would be to delegate these computations to the smartphone. Such a solution has two main drawbacks. On the one hand, delegating computations to an untrusted environment, i.e., the smartphone which could be compromised by spywares, will affect the user's privacy. On the other hand, verifying the received refund token would lead to inefficient transactions at exit turnstiles. 

\end{fullversion} 

\section{Proof of Theorem 2}
\label{appendix:proof2}

\textbf{Sketch of proof.}
The \textbf{completeness} of the protocol follows by inspection. The \textbf{soundness} follows from the extraction property of the underlying proof of knowledge and the unforgeability of the underlying signature scheme. In particular, the extraction property implies that for any prover $P^*$ that convinces $V$ with probability $\epsilon$, there exists an extractor which interacts with $P^*$ and outputs a witness $(k,v,l)$ with probability poly($\epsilon$). Moreover, if we assume that the extractor input consists of two transcripts, i.e., \{$Y, \sum, Com, B, D, c, \tilde{c}, s_1, \tilde{s}_{1},$ $s_2,$ $\tilde{s}_{2}, s_3, \tilde{s}_{3}$\}. The witness can be obtained by computing (all the computations are done mod p): $l = (s_3 - \tilde{s}_{3})/(c - \tilde{c})$ and  $k = (s_1 - \tilde{s}_{1})/(c - \tilde{c})$;  $\nu = (s_2 - \tilde{s}_{2})/(c - \tilde{c})$;

The extractor succeeds when $(c - \tilde{c})$ is invertible in $\mathbb{Z}_p$. If the following check $D = B^y$ holds , this implies that: $D = B^y = B_1^k g^l$. Therefore, $B^y B^k=g^l$. Let us denote by $A_k=B^{1/l}$. Note that we necessarily have $l \neq 0$, otherwise this would imply that the prover knows the secret value $y$ (which would be equal to $-k~mod~p$).

We therefore have $A_k^{y+k}=g$ and consequently that $A_k=g^{1/(y+k)}$. So the prover knows a valid BB-signature on the value $k$. This implies that $k \in \Phi$.

Also note that if $D = B^y$  then this implies that the Prover $P$ only knows one representation of $D$ in the base $B_1$ and $g$. Otherwise this would imply that $P$ knows the private key $y$. Indeed, suppose that $P$ knows two representations of $D$ in the base $B_1$ and $g$. Let us denote by $(k, l)$ and $(\tilde{k}, \tilde{l})$ these two representations. Since $G_1$ is a prime order group and $B_1 \neq 1$ and $g \neq 1$, this implies that $k \neq \tilde{k}$ and $l \neq \tilde{l}$. Since $D = B_1^k g^l = B_1^{\tilde{k}} g^{\tilde{l}}$, this implies that $B_1^{k-\tilde{k}} = g^{l-\tilde{l}}$ and that $B_1^{(k-\tilde{k})/(l-\tilde{l})} = g$. Let us denote by $\tilde{y} = (k-\tilde{k})/(l-\tilde{l})$. Then $D = B^y = B_1^k g^l = B^{-k} B^{-(l\tilde{y})}$. Since $G_1$  is a prime order group this implies that $y = -k - l\tilde{y}$.

If $k \notin \Phi$, then $P^*$ can be directly used to mount a weak-chosen attack against the BB-signature scheme with probability poly($\epsilon$) of succeeding. Thus $\epsilon$ must be negligible.

Finally, to prove \textbf{(honest-verifier) zero-knowledge}, we construct a simulator \textit{Sim} that will simulate all interactions with any (honest verifier) $V^*$.
\begin{enumerate}
\item \textit{Sim} retrieves $Y$ and $\sum$ from $V^*$.
\item \textit{Sim} randomly chooses $k \in \Phi$ and $l \in \mathbb{Z}_p^*$ and computes $B = A_k^l$  , $B_1 = B^{(-1)}$ and $D = B_1^k g^l$.
\item \textit{Sim} randomly chooses $c, s_1, s_2, s_3 \in \mathbb{Z}_p^*$ and computes $Com_1 = g_1^{s_1} h_T^{s_2} Com^{-c}$ and $D_1 = B_1^{s_1} g^{s_3} D^{-c}$.
\item \textit{Sim} outputs $S = \{Com, B, D, Com_1, D_1, c, s_1, s_2,$ $s_3\}$.
\end{enumerate}
Since $G_1$  is a prime order group, then the blinding is perfect in the first two steps and $S$ and $V^*$'s view of the protocol are statistically indistinguishable.

\section{Proof of Theorem 3}
\label{appendix:proof3}

\textbf{Sketch of proof.}
The \textbf{completeness} follows by inspection.

\textbf{Soundness:} roughly speaking, the whole proof $\Pi$ can be divided in three sub-proofs $\Pi_1$, $\Pi_2$ and $\Pi_3$, where:

\noindent $\Pi_1=POK(\alpha,s,r,r_2,r_3,r_5,a: C=g_1^{\alpha s} \times h^\alpha \times B'^r \wedge T'=G^s H^{r_2} \wedge T''=T'^\alpha H^{r_3}=G^{\alpha s} H^{r_5} \wedge C_1=g_T^a \wedge C_2=g_1^s \times h_T^a)$

\noindent $\Pi_2=POK(k,\nu,s,r_2: Com=g_1^k h_T^\nu \wedge T'=G^s H^{r_2} \wedge K=g_t B_k^{-1}=B_k^{s+k})$

\noindent $\Pi_3=POK(k,\nu,l: Com=g_1^k h_T^\nu \wedge D=B_1^k g^l)$

If the verifier accepts the proof $\Pi$ this means that: $D=B^y$ (1) and $C= B_0^\gamma$ (2).

Note that the proofs $\Pi_1$, $\Pi_2$ and $\Pi_3$ are classical variants of Schnorr's proof of knowledge. Using the ``extractors'' of these proofs of knowledge, we can retrieve $k,\alpha,s,r,l$.

From (1) we have: $D=B^y=B_1^k g^l$. This implies that $B^y B^k=g^l$. Let us denote by $A_k=B^{1/l}$. (Note that $l\neq0$, otherwise this would imply that the prover knows the secret value $y$). We therefore have $A_k^{y+k}=g$ and therefore that $A_k=g^{1/(y+k)}$. So the prover knows a valid BB-signature on the value $k$. This implies that $k\in[1..max_{ticket}]$.

From (2) we have that $C= B_0^\gamma=g_1^{\alpha s} \times h^\alpha \times B'^r$. This implies that $B_0^{\gamma+r}=g_1^{\alpha s} \times h^\alpha$. Let us denote by $A=B_0^{1/\alpha}$, this implies that $A^{\gamma+r}=g_1^s \times h$ (Note that $\alpha\neq0$, otherwise this would imply that the prover knows the secret value $\gamma$). So  $A=(g_1^s h)^{1/(\gamma+r)}$. The prover therefore knows a valid permission token ($A, r, s)$. 

In conclusion, the prover knows a permission token $(A, r, s)$ and a value $k$ such that $B_k=g_t^{1/(s+k+1)}$, $k \in[1..max_{ticket}]$ and $E=(C_1,C_2)$ is an El Gamal encryption of $g_1^s$.

\textbf{(Honest-verifier) Zero-Knowledge:} since $\Pi_1$, $\Pi_2$ and $\Pi_3$ are classical ZKPK, we can easily construct simulators $Sim1$, $Sim2$ and $Sim3$ (respectively) for such proofs. From these simulators, it is straightforward to construct a simulator $Sim$ that will simulate all interactions with any (honest) verifier $V^*$. Since $G_1$  is a prime order group, then the blinding would be perfect and the output of  $Sim$ and $V^*$'s view of the protocol would be statistically indistinguishable.

\section{Platform and implementation details}
\label{appendix:param}

\subsection{SIM card details}
We used  Javacard 2.2 for developing all the cryptographic computations into the SIM card. In particular, the SIM card has access to the functions handling elliptic curves operations. The SIM card  handles requests from the validator using the NCF contactless interface.

\subsection{Validator details}
The validator has been developed in Java and Scala. The Java software uses a native library for EC scalar multiplications and pairings. This library depends on libGMP for big integers computations and benefits from its assembly optimizations. Furthermore, computations are distributed between threads (at JVM level) to benefit from the multi-core architecture of the PC. The PC is an Intel(R) Xeon(R) with a E5-1620 CPU with 4 cores running at 3.70GHz under a 64-bit Linux OS. The NFC reader of the validator is an Omnikey 5321 dual interfaces.

\subsection{Curve and Pairing Parameters}
The curve and pairing parameters that we used in the implementation of our m-ticketing system are:

{\noindent\scriptsize{}q = 8243401665430090752057404098378368203946728292799613002

4655912292889294264593 }{\scriptsize \par}

{\noindent\scriptsize{}p = 8243401665430090752057404098378368203918016968090658713

6896645255465309139857 }{\scriptsize \par}

{\noindent\scriptsize{}b = 5 }{\scriptsize \par}

\section{Security proofs}
\label{appendix:proofs}

\subsection{Preliminaries}
\textbf{Discrete Logarithm (DL).} For any multiplicative group $G$ of prime order $p$, the DL assumption states that given a random generator $g \in G$ and a candidate $Y \in G$, it is hard to find an integer $y~mod~p$ such that $Y = g^y$.

\textbf{One-more Discrete Logarithm (OMDL)~\cite{BPS03}.} For any multiplicative group $G$ of prime order $p$, the OMDL assumption states that given a random generator $g \in G$, a challenge oracle that produces a random group element $Y_i$ when queried and a discrete logarithm oracle (with respect to $g$), it is hard to find, after $t$ queries to the challenge oracle (where $t$ is chosen by the adversary) and at most $t - 1$ queries to the discrete logarithm oracle, the discrete logarithms of all $t$ elements $Y_i$.

\textbf{q-Strong Diffie-Hellman I (q-SDH-I).} For any multiplicative group $G_1$ of prime order $p$, the q-SDH-I assumption states that given three random generators ($g_0$, $g_1$, $h$)$\in G_1^3$, a value $W'=g_0^\gamma$, an oracle $\mathcal{O}$ that on input a value $s \in \mathbb{Z}_p$ outputs a pair $(r, A=(g_1^s h)^{1/(\gamma+r)})$ with $r \in \mathbb{Z}_p$ , it is hard to output a new triplet $(r', s', A'=(g_1^{s'} h)^{1/(\gamma+r')})$, with $(r', s') \in \mathbb{Z}_p^2$ such that $s'$ has never been queried to $\mathcal{O}$.

\textbf{Lemma 1.} If the q-SDH assumption holds in $G_1$ then the q-SDH-I assumption holds in $G_1$.

\textbf{Proof.} See~\cite{FPV09} for a proof of this Lemma.

\textbf{Remark:} The triplet $(r',s',A')$ corresponds to a permission token of our m-ticketing protocol. In the sequel, we will call the oracle $\mathcal{O}$, a BB-signature oracle and the value $s'$ the (permission) token secret.

\textbf{Forking Lemma.} We use the Forking Lemma~\cite{PS00} in our proofs, to prove that an adversary $\mathcal{A}$ is not able to produce a new valid m-ticket $Tick_k$ (respectively a Schnorr's signature $\mu$, see Figure~6) unless he knows all the underlying secrets $(a, s, r, A, k)$ (respectively the secret $x_U$).

Using the notation of~\cite{PS00}, if an adversary is able to produce a valid ticket $Tick_k$ $(k, \sigma_1, h, \sigma_2)$ where

\noindent$\sigma_1$=($C$, $C_1$, $C_2$, $B_0$, $T'$, $T''$, $Com$, $B$, $D$, $K$, $C^{'}_1$, $C^{'}_2$, $R'$, $R''$, $C'$, $T^{'}_4$, $Com_1$, $D_1$, $K'$, $L'$, $ch$),

\noindent$h$=$H(C$, $C_1$, $C_2$, $B_0$, $T'$, $T''$, $Com$, $B$, $D$, $K$, $C^{'}_1$, $C^{'}_2$, $R'$, $R''$, $C'$, $T^{'}_4$, $Com_1$, $D_1$, $K'$, $L'$, $ch$) and 

\noindent$\sigma_2$=($s_1$, $s_2$, $s_3$, $\omega_1$, $\omega_2$, $\ldots$, $\omega_{12}$),

\noindent then it can produce two valid m-tickets $(k, \sigma_1, h, \sigma_2)$ and $(k, \sigma_1, h', \sigma_2)$ such that $h \neq h'$.

\subsection{Proof of non-frameability}

\textbf{Sketch of proof.} We assume that the challenger $\mathcal{C}$ in the experiment $\mathbf{Exp}_{\mathcal{A}}^{Nfra}$ $(1^\lambda)$ receives a random instance ($g^{x_1}$, $g^{x_2}$, $\ldots$, $g^{x_l}$) of the one-more DL problem where $g$ is a random generator in $G_1$. $\mathcal{C}$ will run the adversary $\mathcal{A}$ as a subroutine and act as $\mathcal{A}$'s challenger in the non-frameability experiment. Because $\mathcal{C}$ is against the one-more DL assumption, he has access to a DL oracle. $\mathcal{C}$ picks three random values $\xi_1$, $\xi_2$ and $\xi_3$ in $\mathbb{Z}_p$ and computes $g_1=g^{\xi_1}$, $g_U=g^{\xi_2}$ and $G=g_1^{\xi_3}$. The other generators of the m-ticketing system are randomly chosen. $\mathcal{A}$ chooses the keys $tsk$ for the transport authority and $rsk$ for the revocation authority and $\mathcal{C}$ chooses the key $sk_{RP}$  for the Paillier encryption scheme. $\mathcal{C}$ is therefore able to construct the public parameters $pp$ for $\mathcal{A}$ and can answer to its requests in the following way:

\begin{itemize}
\itemsep-0em 
	\item $\mathcal{O}Register_{HU}$ requests: $\mathcal{C}$ answers using his input of the one-more DL problem. $\mathcal{C}$ puts $h_{U_i}$= $(g^{x_i})^{\xi_2}$= $g_U^{x_i}$
	\item $\mathcal{O}Register_{MU}$ requests: $\mathcal{C}$ does not need to do anything.
	\item $\mathcal{O}CorruptUser$ requests: $\mathcal{C}$ uses is DL oracle to give the corresponding $x_i$ to the adversary.
	\item $\mathcal{O}TokenRequest_U$ requests: $\mathcal{C}$ first uses his input of the one-more DL problem to compute the commitment $Com$: $\mathcal{C}$ puts $Com$= $(g^{x_j})^{\xi_1}$= $g_1^{x_j}$. If the protocol does not abort then we put $s$=$x_j$+$s_2~mod~p$ and $c = g_1^s$, where $x_j$ is unknown to $\mathcal{C}$ and $s_2$ is provided by $\mathcal{A}$. In the random oracle model, $\mathcal{C}$ is able to perfectly simulate the proof of knowledge $\Pi_1$ as well as the Schnorr's signature $\mu$.
	\item $\mathcal{O}GenTicket(ID_U, \emph{view})$ requests: All the values involved in the computation of an m-ticket $Tick_k$ can be easily simulated by $\mathcal{C}$ except $T'$ and $B_k$. To compute $T'= G^s H^{r_2}$ where $r_2$ is a random value chosen by $\mathcal{C}$, it proceeds as follows: $T'$ = $(Com \times g_1^{s_2})^{\xi_3} H^{r_2}$ = $G^s H^{r_2}$. As $\mathcal{C}$ does not know the value $s$ it cannot compute or simulate the value $B_k=g_t^{1/(s+k+1)}$. It will therefore choose a random value $R$ and define $B_k$ as $R$. In the random oracle model, $\mathcal{C}$ can simulate the proof of knowledge $\Pi$ using standard techniques. The simulation is not perfect since we replace the value $B_k$ by a random value $R$. However $\mathcal{A}$ will not detect this change unless he can solve the q-DDHI problem.
	\item $\mathcal{O}ReportTicket(ID_U, \emph{view})$ requests: $\mathcal{C}$ proceeds as in an $\mathcal{O}GenTicket$ request for each unused m-ticket.
\end{itemize}
Now, we assume that the adversary $\mathcal{A}$ manages to produce a valid m-ticket $Tick_k$ such that it breaks the non-frameability of our m-ticketing protocol and it mades $t$ requests to the $\mathcal{O}CorruptUser$ oracle. This means that this m-ticket has not been obtained on a $\mathcal{O}GenTicket$ query and the $\texttt{IdentUser}$ algorithm on $Tick_k$ outputs an identifier $ID_U$ which is in $\mathcal{HU}$ along with a Schnorr's signature $\mu$ that proves that this m-ticket comes from a permission token obtained by this user\footnote{\scriptsize We would like to emphasize that since the output of $\texttt{IdentUser}$ can be publicly verifiable, a wrong $\texttt{IdentUser}$ procedure is statistically negligible (even for a powerful adversary).} .

It follows from the Forking Lemma that if $\mathcal{A}$ is able to produce a valid m-ticket $Tick_k$ $(k, \sigma_1, h, \sigma_2)$ where

\noindent$\sigma_1$=($C$, $C_1$, $C_2$, $B_0$, $T'$, $T''$, $Com$, $B$, $D$, $K$, $C^{'}_1$, $C^{'}_2$, $R'$, $R''$, $C'$, $T^{'}_4$, $Com_1$, $D_1$, $K'$, $L'$, $ch$),

\noindent$h$=$H(C$, $C_1$, $C_2$, $B_0$, $T'$, $T''$, $Com$, $B$, $D$, $K$, $C^{'}_1$, $C^{'}_2$, $R'$, $R''$, $C'$, $T^{'}_4$, $Com_1$, $D_1$, $K'$, $L'$, $ch$) and 

\noindent$\sigma_2$=($s_1$, $s_2$, $s_3$, $\omega_1$, $\omega_2$, $\ldots$, $\omega_{12}$) then it can produce two valid m-tickets $(k, \sigma_1, h, \sigma_2)$ and $(k, \sigma_1, h', \sigma_2)$ such that $h \neq h'$. Using the technique of replay and the soundness property of the proof $\Pi$, one is able to extract all the secret values ($a$, $s$, $r$, $A$, $k$).

\textbf{First case:} the value $s$ corresponds to a permission token obtained during an $\mathcal{O}TokenRequest_U$ on $ID_{U_i}$ (i.e., $g_1^s$ is equal to the value $c$ produced during such a request). $\mathcal{C}$ outputs the $t$ values $x_i$ that comes from the requests to the DL oracle plus the value $x_j=s- s_2~mod~p$ and so breaks the one-more DL assumption.  

\textbf{Second case:} the value $s$ does not correspond to a permission token obtained during an $\mathcal{O}TokenRequest_U$ on $ID_{U_i}$ (meaning that all the values $c$ generated during such requests are different from $g_1^s$). We know by the definition of the experiment that no $\mathcal{O}CorruptUser$ oracle query (and consequently no DL oracle query) has been made on this identity. Therefore the public key $h_{U_i}$ corresponding to $ID_{U_i}$ is in the one-more DL problem input. It follows from the Forking Lemma that if $\mathcal{A}$ is sufficiently efficient to produce such a signature $\mu$, then there exists an algorithm A' which can produce two Schnorr's signatures with non-negligible probability. Using the techniques of replay and soundness, $\mathcal{C}$ is able to extract the private key $x_U$ used to generate the signature $\mu$. $\mathcal{C}$ outputs the t values $x_i$, coming from the requests to the DL oracle, plus the value $x_{U_i}$  and so breaks the one-more DL assumption.

We prove the non-frameability under the q-DDHI and one-more discrete logarithm assumptions~\cite{BPS03}. We use in fact the OMDL assumption to get a better reduction, but the proof can also be done under the discrete logarithm assumption. As the q-DDHI assumption implies the DL one, we can conclude that our m-ticketing protocol is non-frameable, in the random oracle model, under the q-DDHI assumption.

\subsection{Proof of unforgeability}

\textbf{Sketch of proof.} Let $\mathcal{A}$ be an adversary who breaks the unforgeability requirement of our m-ticketing protocol with non-negligible probability. We will construct an algorithm $\mathcal{B}$, using $\mathcal{A}$ as an oracle, which breaks the q-SDH-I assumption. $\mathcal{B}$ receives on input from its challenger ($G_1$, $g_0$, $g_1$, $g_U$ $h$, $W'=g_0^\gamma$) the public parameters of the q-SDH-I challenge and has access to a BB-signature oracle. The other generators of the m-ticketing system are randomly chosen.

$\mathcal{B}$ also chooses the keys for the Paillier encryption scheme. It sends $W'$ and the public key of the Paillier scheme to $\mathcal{A}$. The private and public keys of the El Gamal cryptosystem can be chosen either by $\mathcal{A}$ or $\mathcal{B}$. $\mathcal{B}$ is therefore able to construct the public parameters $pp$ for $\mathcal{A}$ and can answer to its requests as follow:

\begin{itemize}
\itemsep-0em 
	\item  $\mathcal{O}Register_{HU}$ requests: $\mathcal{B}$ randomly chooses $x_U \in \mathbb{Z}_p$ and computes $h_U= g_U^{x_U}$.
	\item $\mathcal{O}Register_{MU}$  requests: $\mathcal{B}$ does not need to do anything.
	\item $\mathcal{O}CorruptUser$ requests: $\mathcal{B}$ gives $x_U$  to $\mathcal{A}$.
	\item $\mathcal{O}TokenRequest_T$ requests: $\mathcal{B}$ plays as follows: $\mathcal{B}$ first receives a Paillier encryption $C_0$. It decrypts it (recall that $\mathcal{B}$ chose the private key for the Paillier encryption scheme) and retrieve the corresponding plaintext $s_1$. It then queries the BB-signature oracle on $s$ where $s=s_1+s_2$ and $s_2$ is randomly chosen by $\mathcal{B}$. The BB-signature oracle sends back a pair $(r, A=(g_1^s h)^{1/(y+r)})$ with $r \in \mathbb{Z}_p$ , and $\mathcal{B}$ transmits it to $\mathcal{A}$ along with the value $s_2$. So $\mathcal{B}$ perfectly simulates the $\mathcal{O}TokenRequest_T$ for $\mathcal{A}$.
	\item $\mathcal{O}GenTicket(ID_U, \emph{view})$ requests: since $\mathcal{B}$ knows the values of all the tokens $(A, r, s)$ that have been issued during $\mathcal{O}TokenRequest_T$ requests, it can easily answer to $\mathcal{O}GenTicket$ queries. The simulation of this oracle is perfect.
	\item $\mathcal{O}ReportTicket(ID_U, \emph{view})$ requests: $\mathcal{B}$ proceeds as in an $\mathcal{O}GenTicket$ request for each unused m-ticket.

\end{itemize}
We differentiate two types of adversaries:
\begin{itemize}
\itemsep-0em 
	\item \textbf{Type-1 Forger}: an adversary that outputs a valid m-ticket $Tick_k$ which cannot be linked to a registered user (corrupted or not).
	\item \textbf{Type-2 Forger}: an adversary that outputs more valid m-tickets than he obtained.
\end{itemize}
We show that both Forgers can be used to solve the q-SDH-I problem. However, the reduction works differently for each Forger type. Therefore, initially $\mathcal{B}$ chooses a random bit $c_{mode} \in {1, 2}$ that indicates its guess for the type of forgery that $\mathcal{B}$ will output.

\textbf{If $\mathbf{c_{mode}}$ = 1}: Eventually, $\mathcal{B}$ outputs, with non-negligible probability, a valid m-ticket $Tick_{k'} = (B_{k'}, E', \Pi')$ such that the algorithm $\texttt{IdentUser}$, on the input $Tick_{k'}$, returns an unknown identifier $ID$ (i.e., does not appear in $DB_{REG}$).

\textbf{First case}: $c=g_1^{s'}$ (the plaintext encrypted in $E'$) has been queried during an $\mathcal{O}TokenRequest_T$ request but no corresponding signature $\mu$ has been produced. This means that the adversary did not receive the value $s_2$ (where $s'= {s'}_1+s_2$) from the $\mathcal{O}TokenRequest_T$ oracle. We know from the Forking Lemma and the proofs $\Pi$ and $\Pi_1$ that $\mathcal{A}$ necessarily knows $s'$ and $s_2$. Since only $g_1^{s_2}$ has been revealed by this oracle during the $\mathcal{O}TokenRequest_T$, $\mathcal{A}$ could be used to extract Discrete Logarithms. Therefore under the DL assumption, this first case could only occur with negligible probability.

\textbf{Second case}: $E'$ is an encryption of a commitment $c=g_1^{s'}$ of a value $s'$ that has not been queried during a $\mathcal{O}TokenRequest_T$ request. Therefore $s'$ has not been queried to the BB-signature oracle either. Using the Forking Lemma and the soundness property of the proof $\Pi_1$, $\mathcal{B}$ is able to extract with non-negligible probability the secrets $(a',s',r',A',k')$ underlying the m-ticket $Tick_{k'}$. $\mathcal{B}$ outputs the triplet $(r',s',A')$ to its challenger of the q-SDH-I assumption and therefore breaks it.

\textbf{If $\mathbf{c_{mode}}$ = 2}: Eventually, $\mathcal{A}$ outputs, with non-negligible probability $\xi$, $L =N \times max_{ticket}+1$ valid m-tickets \footnote{\scriptsize Without loss of generality, we do not make any distinction between a m-ticket and an unused m-ticket}  $\{Tick_{k_j}^j \}_{j=1}^{j=L}$, where $N$ is the number of calls to the $\mathcal{O}TokenRequest_T$ oracle, $max_{ticket}$ is the number of authorized m-tickets by token and $Tick_{k_j}^j = (B_{k_j}, E, \Pi)$. Let us denote by ($s_1$, $s_2$, $\ldots$, $s_N$) the $N$ token secrets submitted to the BB-signature oracle. W.l.o.g, we may assume that all these values are different (recall that a token secret $s$ is jointly computed by $\mathcal{A}$ and $\mathcal{B}$). We therefore have $N \times max_{ticket}$ distinct token pairs $(s_i,k)$ with $i \in \{1,\ldots,N\}$ and $k \in \{1,\ldots, max_{ticket}\}$. Let $\Gamma$ be the set containing all these token pairs and $Tick_j^i$ the m-ticket corresponding to the token pair $(s_i,j)$.

Among the $L$ m-tickets output by $\mathcal{A}$, there is at least one m-ticket $Tick_{k*}$ for which the corresponding token pair \textbf{($s*$, $k*$) does not belong to $\Gamma$}. Otherwise this would mean that two m-tickets among these $L$ m-tickets, e.g., $Tick_{k_1}$ and $Tick_{k_2}$, have the same token pair (since $L > N \times max_{ticket}$). Let us denote by ($s_1^*$, $k_1$) (respectively ($s_2^*$, $k_2$)) the token pair corresponding to $Tick_{k_1}$ (respectively $Tick_{k_2}$). Therefore the serial number $B_{k_1}^*$ of $Tick_{k_1}$ would be equal to $B_{k_2}^*$ the one of $Tick_{k_2}$: $B_{k_1}^*$= $g_t^{1/(s_1^*+k_1+1)}$= $g_t^{1/(s_2^*+k_2+1)}$ = $B_{k_2}^*$. This case cannot occur since all duplicates (i.e. m-tickets which have the same serial numbers) are discarded in the experiment $\mathbf{Exp}_{\mathcal{A}}^{unforg}(1^\lambda)$.

Suppose now that \textbf{$s* \in \{s_1,s_2,\ldots,s_N\}$}. Since $(s*,k*) \notin \Gamma$ this implies that $k* \notin \{1,\ldots,max_{ticket}\}$. Such a case will happen with negligible probability under the q-SDH assumption (see Theorem 2). Therefore $s* \notin \{s_1, s_2,\ldots,s_N\}$ and consequently has not been queried to the BB-signature oracle ($\mathcal{A}$ is in fact a Type-1 Forger).  $\mathcal{B}$ then picks a random m-ticket $Tick_{k'}$ among the $L$ m-tickets output by $\mathcal{A}$. With probability $1/L$, it has chosen $Tick_{k*}$. Using the Forking Lemma and the soundness property of the proof $\Pi$, $\mathcal{B}$ is able to extract with non-negligible probability the secrets $(a',s',r',A',k')$ underlying the m-ticket $Tick_{k'}$. $\mathcal{B}$ outputs the triplet $(r',s',A')$ and therefore breaks the q-SDH-I assumption with non-negligible probability $\xi/L$.

To complete the proof, we need to clarify why we discard duplicates in $\mathbf{Exp}_{\mathcal{A}}^{unforg}(1^\lambda)$. We consider that $Tick_k$ = $(B_k, E, \Pi)$  and $Tick_{k'}$ = $(B_{k'} , E', \Pi')$  are duplicates if their serial numbers are equal. Let us denote by $(s, k)$ (respectively $(s', k')$) the token pair corresponding to the ticket $Tick_k$ (respectively $Tick_{k'}$). Since $B_k$ = $B_{k'}$, we have $s + k$ = $s' + k'~mod~p$.

\textbf{First case}: $(s, k)$ = $(s', k')$: This implies that $Tick_k$ and $Tick_{k'}$ are in fact the same tickets. We can therefore discard one of them.

\textbf{Second case}: $(s, k)$ $\neq$ $(s', k')$:

\textit{Case 2.1}: $s$ or $s'$ $\notin$ $\{s_1, s_2,\ldots, s_N\}$. This implies that $\mathcal{A}$ is a Type-1 Forger. Under the q-SDH-I assumption, this case will occur with negligible probability.

\textit{Case 2.2}: $s$ and $s'$ $\in$ $\{s_1, s_2,\ldots, s_N\}$. This implies that $k$ and $k'$ $\in$ $\{1,\ldots, max_{ticket}\}$. Otherwise $\mathcal{A}$ could be used to break the q-SDH assumption (see the proof of Theorem 2). Since $s$ and $s'$ have been randomly chosen (they are jointly computed by $\mathcal{A}$ and $\mathcal{B}$), the probability that $s + k$ = $s' + k'~mod~p$ is negligible. Therefore Case 2.2. will occur with negligible probability either.

Consequently, under the q-SDH assumption, only the first case could occur and we only discard tickets that come from the same token secret. 

\begin{fullversion}

\subsection{Proof of unlinkability}

\textbf{Sketch of proof.} We prove the \textbf{unlinkability} under a slightly weaker model than the one presented in Section~\ref{unlink}. Indeed, we consider a slightly different experiment in which the adversary cannot query the revocation oracle $\mathcal{O}IdentUser_T$. We rename this new requirement \textbf{Unlinkability*}. We would like however to emphasize that the access to revocation functionalities will likely be carefully controlled in a real deployment of an m-ticketing system. Therefore, unlinkability* is a reasonable model to consider.

Anyway, in order to satisfy the original unlinkability requirement, we just need to replace the ElGamal encryption scheme, which only satisfies IND-CPA security, by an IND-CCA2 encryption scheme. It is well-known that by double encrypting the same message under an IND-CPA scheme and using a \emph{simulation sound proof of equality} of plaintexts, we obtain an IND-CCA2 scheme. Therefore, by using the double ElGamal encryption scheme, which was proved IND-CCA2 in~\cite{FP01}, our m-ticketing protocol would satisfy the original unlinkability requirement.

Let $\mathcal{A}$ be an adversary who breaks the unlinkability requirement of our m-ticketing protocol. W.l.o.g. we will consider that a query to the $\mathcal{O}ReportTicket$ oracle on ($ID_U$, \emph{view}) for the report of $j$ unused m-tickets is equivalent to $j$ queries to the $\mathcal{O}GenTicket$ on ($ID_U$, \emph{view}).

Let $m$ = $max_{ticket}$. We will say that an adversary $\mathcal{A}$ against our unlinkability experiment $\mathbf{Exp}_{\mathcal{A}}^{unlink-b}(1^\lambda)$ is a Type-($i$, $j$) distinguisher, with $i \leq m-1$ and $j \leq m-1$, if $\mathcal{A}$ after having received its challenge (from its $\mathbf{Exp}_{\mathcal{A}}^{unlink-b}(1^\lambda)$-challenger) makes at most $i$ queries to the $\mathcal{O}GenTicket$ oracle on ($i_0 $, $\emph{view}_0$) and $j$ queries to the $\mathcal{O}GenTicket$ oracle on ($i_1$, $\emph{view}_1$). We can remark that a Type-($i$, $j$) distinguisher, with $i \leq m-1$ and $j \leq m-1$, is also a Type-($m-1$, $m-1$) distinguisher. We may therefore, without loss of generality, restrict our attention to Type-($m-1$, $m-1$) distinguishers. In the sequel, we will thus assume that $\mathcal{A}$ is such an adversary. More precisely, we will suppose that after receiving $Tick_{k_b}$ and $Tick_{k_{1-b}}$, $\mathcal{A}$ arbitrarily queries the $\mathcal{O}Register_{HU}$, $\mathcal{O}Register_{MU}$, $\mathcal{O}CorruptUser$, $\mathcal{O}IdentTicket_T$ and $\mathcal{O}TokenRequest_U$ oracles and then queries the $\mathcal{O}ReportTicket$ oracle on ($i_0 $, $\emph{view}_0$) and ($i_1$, $\emph{view}_1$).

We give a security proof as a sequence of games, using Shoup's methodology~\cite{Shoup04}. We will only give a rather high-level description of the initial game (Game 0) and brief descriptions of the modifications between successive games.

\textbf{Game 0}: This is the original attack game with respect to a given efficient adversary $\mathcal{A}$.

The challenger $\mathcal{C}$ randomly chooses $g$, $g_0$, $g_1$, $g_t$, $g_T$, $g_U$, $h$, $G$, $H$ nine generators of $G_1$ and $g_2$, $g_3$ two generators of $G_2$. It also chooses the keys for the Paillier encryption scheme as well as the ones for the transport authority and revocation authorities. $\mathcal{C}$ sends $gpk$ and $tsk$ to $\mathcal{A}$. $\mathcal{C}$ answers to $\mathcal{A}$'s requests as follows:

\begin{itemize}
	\item $\mathcal{O}Register_{HU}$ requests: $\mathcal{C}$ randomly chooses $x_U \in \mathbb{Z}_p$ and computes $h_U$ = $g_U^{x_U}$.
	\item $\mathcal{O}Register_{MU}$ requests: $\mathcal{C}$ does not need to do anything.
	\item $\mathcal{O}CorruptUser$ requests: $\mathcal{C}$ gives $x_U$ to $\mathcal{A}$.
	\item $\mathcal{O}TokenRequest_U$ requests: $\mathcal{C}$ chooses a random value $s_1$ to obtain a valid permission token $(A, r, s)$ and uses $x_U$ to generate the signature $\mu$.
	\item $\mathcal{O}GenTicket(ID_U, \emph{view})$ requests: $\mathcal{C}$ uses its permission token $(A, r, s)$ and a fresh index $k$ that has not been used in a previous query of $\mathcal{O}GenTicket$ on $ID_U$ and \emph{view} and computes a ticket $Tick_k$ = $(B_k, E, \Pi)$. 
	\item $\mathcal{O}IdentTicket_T(ID_U,  \emph{view})$ requests: $\mathcal{C}$ computes the m-tickets $Tick_k$ based on the secret $s$ corresponding to the user identifier $ID_U$ and  gives them to $\mathcal{A}$.
	\item $\mathcal{O}ReportTicket(ID_U, \emph{view})$ requests: $\mathcal{C}$ proceeds as in a $\mathcal{O}GenTicket$ request for each unused m-ticket.
	
\end{itemize}

The adversary chooses two honest users $i_0$ and $i_1$ and two indexes $k_0$ and $k_1$. $\mathcal{C}$ runs the protocol $\texttt{TokenRequest}$ with $i_0$ and $i_1$ and outputs the corresponding views, $\emph{view}_0$ and $\emph{view}_1$, to $\mathcal{A}$. It then generates two valid m-tickets: $Tick_{k_b}$ = $(B_{k_b}, E_b, \Pi_b)$ for $i_0$ and $Tick_{k_{1-b}}$ = $(B_{k_{1-b}}, E_{1-b}, \Pi_{1-b})$ for $i_1$ and send them to $\mathcal{A}$. The goal of $\mathcal{A}$ is to guess the value of $b$. After having received its challenge, $\mathcal{A}$ can queries the $\mathcal{O}Register_{HU}$, $\mathcal{O}Register_{MU}$, $\mathcal{O}CorruptUser$, $\mathcal{O}TokenRequest_U$, $\mathcal{O}GenTicket$, $\mathcal{O}IdentTicket_T$ and $\mathcal{O}ReportTicket$ oracles but with the following restrictions: it cannot query the $\mathcal{O}CorruptUser$ oracle on $i_0$ or $i_1$ or the $\mathcal{O}IdentTicket_T$ oracle  on ($i_0 $, $\emph{view}_0$) or ($i_1$, $\emph{view}_1$) and cannot query the $\mathcal{O}Report$ $Ticket$ oracle on ($i_0 $, $\emph{view}_0$) or ($i_1$, $\emph{view}_1$) if both users did not validate the same number of m-tickets (otherwise it can easily win this game).

At the end of the game, the adversary outputs a bit $b'$, its guess. Let $S_0$ be the event that $b = b'$ in this game and $S_i$ the event that $b = b'$ in game i. We have:
$|Pr[S_0]-1/2|$ = $\mathbf{Adv}_{\mathcal{A}}^{unlink-b}(1^\lambda)$ = $|Pr[\mathbf{Exp}_{\mathcal{A}}^{unlink-b}(1^\lambda) = b]- 1/2|$

Let $\{Tick^i_{k_b^i} = (B^i_{k_b^i}, E^i_{k_b^i}, \Pi^i_{k_b^i})\} _{i=1}^{i=m-1}$ denote the $m-1$ unused (reported) m-tickets of $i_0$ and $\{Tick^i_{k_{1-b}^i}$ = ($B^i_{k_{1-b}^i}$, $E^i_{k_{1-b}^i}$, $\Pi^i_{k_{1-b}^i})\} _{i=1}^{i=m-1}$ the ones of $i_1$, where the $k_b^i$'s and the $k_{1-b}^i$'s belongs to $\{1,\ldots, m\}$.

\begin{itemize}
	\item \textbf{Game(0,b)}: This is the same game as \textbf{Game 0} except that we replace the El Gamal ciphertext $E_b$ by an encryption of a random value and then simulate the proof $\Pi_b$. Such a simulated proof can easily be done in the random oracle model using standard techniques. Under the DDH assumption, $\mathcal{A}$ cannot detect this change. Indeed, we can easily construct a DDH distinguisher $\mathcal{D}$ with DDH-advantage satisfying: \\ $|Pr[S_0]- Pr[S_{(0,b)}]|$ $\leq$ $\mathbf{Adv}_{\mathcal{D}}^{DDH}(1^\lambda)$ (1) \\ where $\mathbf{Adv}_{\mathcal{D}}^{DDH}(1^\lambda)$ is the DDH-advantage of $\mathcal{D}$ in solving the DDH problem. In the sequel, we will use the simplified notation $\mathbf{Adv}_{DDH}$ (respectively $\mathbf{Adv}_{q-DDHI}$) to denote the DDH-advantage (respectively the q-DDHI advantage) of some efficient algorithm in solving the DDH (respectively q-DDHI) problem.
	\item \textbf{Game(0,1-b)}: This is the same game as \textbf{Game(0,b)} except that we replace the El Gamal ciphertext $E_{1-b}$ by an encryption of a random value and then simulate the proof $\Pi_{1-b}$. Such a simulated proof can easily be done in the random oracle model using standard techniques. Under the DDH assumption, $\mathcal{A}$ cannot detect this change. Indeed, we can easily construct a DDH distinguisher D with DDH-advantage satisfying: \\ $|Pr[S_{(0,b)}]- Pr[S_{(0,1-b)}]|$ $\leq$ $\mathbf{Adv}_{DDH}$ (2)	
	\item \textbf{Game(0,0,b)}: This is the same game as \textbf{Game(0,1-b)} except that we replace the serial number $B_{k_b}$ by a random value and then simulate the proof $\Pi_b$. Such a simulated proof can easily be done in the random oracle model using standard techniques. Under the q-DDHI assumption, $\mathcal{A}$ cannot detect this change. Indeed, we can easily construct a q-DDHI distinguisher $\mathcal{D}$ with q DDHI-advantage satisfying: \\ $|Pr[S_{(0, 1-b)}]- Pr[S_{(0,0,b)}]|$ $\leq$ $\mathbf{Adv}_{q-DDHI}$ (3)
	\item 	\textbf{Game(0,0,1-b)}: This is the same game as \textbf{Game(0,0,b)} except that we replace the serial number $B_{k_{1-b}}$ by a random value and then simulate the proof $\Pi_{1-b}$. Such a simulated proof can easily be done in the random oracle model using standard techniques. Under the q-DDHI assumption, $\mathcal{A}$ cannot detect this change. Indeed, we can easily construct a q-DDHI distinguisher $\mathcal{D}$ with q DDHI-advantage satisfying: \\ $|Pr[S_{(0,0,b)}]- Pr[S_{(0,0,1-b)}]|$ $\leq$ $\mathbf{Adv}_{q-DDHI}$ (4)	
\end{itemize}

Let \textbf{Game 1} be the same game as \textbf{Game(0,0,1-b)}

Combining (1), (2), (3) and (4), we obtain:\\ 

$|Pr[S_0]- Pr[S_1]|$ $\leq$ $2 \mathbf{Adv}_{DDH}$ + $2 \mathbf{Adv}_{q-DDHI}$ \\

We then proceed similarly for each unused m-ticket of $i_0$ and $i_1$: $\{Tick^i_{k_b^i} = (B^i_{k_b^i}, E^i_{k_b^i}, \Pi^i_{k_b^i})\} _{i=1}^{i=m-1}$ and $\{Tick^i_{k_{1-b}^i}$ = ($B^i_{k_{1-b}^i}$, $E^i_{k_{1-b}^i}$, $\Pi^i_{k_{1-b}^i})\} _{i=1}^{i=m-1}$. For $i=1$ to $m-1$, we define the following game.  

\begin{itemize}
	\item \textbf{Game(i,b)}: This is the same game as \textbf{Game i} = \textbf{Game(i-1,0,1-b)} except that we replace the El Gamal ciphertext $E_b^i$ by an encryption of a random value and then simulate the proof $\Pi_b^i$.
	\item \textbf{Game(i,1-b)}: This is the same game as \textbf{Game(i,b)} except that we replace the El Gamal ciphertext $E_{1-b}^i$ by an encryption of a random value and then simulate the proof $\Pi_{1-b}^i$.
	\item \textbf{Game(i,0,b)}: This is the same game as \textbf{Game(i,1-b)} except that we replace the serial number $B_{k_b^i}^i$ by a random value and then simulate the proof $\Pi_{b}^i$.
	\item \textbf{Game(i,0,1-b)}: This is the same game as \textbf{Game i,0,b)} except that we replace the serial number $B_{k_{1-b}^i}^i$ by a random value and then simulate the proof $\Pi_{1-b}^i$.
\end{itemize}

Let \textbf{Game i + 1} be the same game as \textbf{Game(i,0,1-b)}. We obtain as above: $|Pr[S_{i+1}]- Pr[S_{i}]|$ $\leq$ $2 \mathbf{Adv}_{DDH}$ + $2 \mathbf{Adv}_{q-DDHI}$

Using our notation, \textbf{Game m} is the same game as \textbf{Game(m-1,0,1-b)}. In the latter game, the unlinkability-challenger gives no information (in a strong information theoretic sense) to $\mathcal{A}$ about the bit $b$ (since all the El Gamal ciphertexts have been replaced by encryptions of random values and all serial numbers have been replaced by random values). Therefore we have: $Pr[S_m] = 1/2$.

We can now give an upper bound for $\mathbf{Adv}_{\mathcal{A}}^{unlink-b}(1^\lambda)$ : \\

$\mathbf{Adv}_{\mathcal{A}}^{unlink-b}(1^\lambda)$ = $|Pr[\mathbf{Exp}_{\mathcal{A}}^{unlink-b}(1^\lambda) = b]- 1/2|$ =  $|Pr[S_0]- Pr[S_m]|$\\

We have:\\

$|Pr[S_0]- Pr[S_m]|$ $\leq$ $\Sigma_{j=0}^{j=m-1} |Pr[S_{j}]- Pr[S_{j+1}]|$ $\leq$ $2m \times (\mathbf{Adv}_{DDH}$ + $\mathbf{Adv}_{q-DDHI})$ \\

Therefore under the DDH and q-DDHI assumptions, the advantage $\mathbf{Adv}_{\mathcal{A}}^{unlink-b}(1^\lambda)$  of a Type-($m-1$, $m-1$) distinguisher is negligible (and consequently the one of a Type-($i$, $j$) distinguisher, with $i \leq m-1$ and $j \leq m-1$, is also negligible).

We can then conclude that our proposed m-ticketing protocol satisfies the unlinkability* requirement, in the random oracle model, under the DDH and q-DDHI assumptions.

\end{fullversion}

\end{document}